\documentclass[12pt]{article}

\usepackage{amsmath,amsfonts,amssymb,latexsym}
\def\mco{\multicolumn}

\numberwithin{equation}{section}
\def\be{\begin{equation}}
\def\ee{\end{equation}}
\def\bq{\begin{eqnarray}}
\def\eq{\end{eqnarray}}
\def\beq{\begin{eqnarray*}}
\def\eeq{\end{eqnarray*}}

\def\a{\alpha}

\begin{document}
\begin{titlepage}
\begin{flushright}
{\tt CERN-PH-TH/2012-152}\\
\end{flushright}

\vspace{1cm}

\begin{center}
{\huge Brane singularities with mixtures in the bulk}

\vspace{1cm}
{\large Ignatios Antoniadis$^{1,*,3}$, Spiros Cotsakis$^{2,\dagger}$, Ifigeneia Klaoudatou$^{2,\ddagger}$}\\

\vspace{0.5cm}

$^1$ {\normalsize {\em Department of Physics, CERN - Theory Division}}\\
{\normalsize {\em CH--1211 Geneva 23, Switzerland}}\\

\vspace{2mm}

$^2$ {\normalsize {\em Research Group of Geometry, Dynamical Systems
and Cosmology}}\\ {\normalsize {\em Department of Information and
Communication Systems Engineering}}\\ {\normalsize {\em University
of the Aegean}}\\ {\normalsize {\em Karlovassi 83 200, Samos,
Greece}}\\
\vspace{2mm} {\normalsize {\em E-mails:}
$^*$\texttt{ignatios.antoniadis@cern.ch},
$^\dagger$\texttt{skot@aegean.gr},
$^\ddagger$\texttt{iklaoud@aegean.gr}}
\end{center}

\vspace{0.5cm}

\begin{abstract}
\noindent
By extending previous analysis of the authors, a systematic study of the
singularity structure and possible asymptotic behaviors of five-dimensional
braneworld solutions is performed in the case where the bulk is a mixture of
an analog of perfect fluid (with a density and pressure depending on the extra coordinate) and a massless scalar field. The two bulk components interact by exchanging
energy so that the total energy is conserved. In a particular range of the interaction parameters, we find flat brane general solutions avoiding the singularity at finite distance
from the brane, in the same region of the equation of state constant parameter
$\gamma=P/\rho$ that we found previously in the absence of the bulk scalar field
$(-1<\gamma<-1/2)$.
\end{abstract}
%
\begin{center}
{\line(5,0){280}}
\end{center}

$^3${\small On leave from {\em CPHT (UMR CNRS 7644) Ecole
Polytechnique, F-91128 Palaiseau, France.}}

\end{titlepage}
\section{Introduction}

In our previous works~\cite{ack_proc,ack}, we started a systematic study of the
singularity structure and possible asymptotic behaviors of five-dimensional
braneworld solutions, by parameterizing the bulk field content with an analog
of perfect fluid satisfying the equation of state $P=\gamma\rho$,
where the `pressure' $P$ and the `density' $\rho$ depend only on the extra dimension $Y$ and $\gamma$ is a constant parameter.
Our motivation was  based on the idea of the so-called self-tuning mechanism
for the cosmological constant~\cite{nima,silver}, aiming to examine in a model
independent way the possibility of avoiding singularities in the bulk at a finite
distance from the brane position. We had found three regions of $\gamma$ leading
to qualitatively different behavior:
\begin{itemize}
\item The region $\gamma>-1/2$ is very similar to the case of a massless bulk
scalar field.
Indeed, the existence of a singularity at a finite distance is unavoidable in
all solutions with a flat brane, in agreement with earlier works that made similar
investigations in different models, using other methods~\cite{Gubser,Forste}.
Moreover, we have shown that 
the singularity can be avoided ({\em e.g.} moved at infinite distance)
when the brane becomes curved, either positively or negatively.
Thus, requiring absence of singularity brings back the cosmological constant problem,
since the brane curvature depends on its tension that receives quartic
divergent quantum corrections.
\item In the region $-1<\gamma<-1/2$, the curved brane solution becomes singular
while the flat brane is regular. Thus, this region seems to avoid the main
obstruction of the self-tuning proposal: any value of the brane tension is
absorbed in the solution and the brane remains flat.\footnote{Of course, the main
question is then whether there is a consistent field theory realization of such a
fluid producing naturally an effective equation of state of this type~\cite{Forste2}.}
\item In the region $\gamma<-1$, corresponding to  the analog of a phantom
equation of state,
the brane can be ripped apart in as much the same way as in a big rip
singularity. This happens only in the flat case, while curved brane solutions
develop `standard' collapse singularities. No regular solution was found in this
region.
\end{itemize}
Moreover, we have shown that all possible singularities at a finite
distance from the original position of the brane can be classified
in three main classes which we coined collapse type I, collapse type
II, and big rip singularities, respectively.

The collapse types I and II met in the asymptotic evolution are both
characterized, as their name suggests, by the vanishing of the warp
factor. Their differences can be traced in the behavior of the
derivative of the warp factor and density of the matter component.
In the collapse type I class for instance, the derivative of the
warp factor becomes infinite whereas in the collapse type II class
it remains finite. The density of the bulk matter, on the other
hand, is necessarily divergent asymptotically in the collapse type I
class, whereas in the collapse type II class it may approach a
constant or even vanish.

An interesting aspect of these two behaviors is that the types of
singularity which become asymptotically feasible depend in the
first place, on the type of bulk matter: while a massless scalar field,
which may be regarded as a fluid with $\gamma =1$,
allows the development of only a collapse type I singularity, a
perfect fluid allows in addition the emergence of
singularities covering the whole variety of the collapse type II
class as well as big rip singularities. The latter are
singularities characterized by the divergence of the warp factor,
its derivative and the matter density of the fluid, and arise only
when the parameter $\gamma$ is less than $-1$. In addition, the
possible types are determined by the spatial geometry of the
brane: A flat brane allows the development of all the different
types of finite-distance singularity whereas a curved brane
permits exclusively the formation of collapse type II
singularities.

In this paper we extend our previous work by examining the case of a bulk filled with a
mixture of fluid and massless scalar field. We let these two bulk entities either interact with
each other or simply coexist independently in the bulk. In the latter
case, we show that all previous types of singularity are still possible. However,
the nature of each type is now enriched with the behavior of both bulk
components. The collapse type I singularity, for example, is characterized by the
divergence of the density of either one or even both of the two components, whereas,
the big rip singularity is characterized by the divergence only of the density of
the fluid while the density of the scalar field vanishes asymptotically. These two types of
singularity arise in general solutions. The collapse type II singularity on the other hand, arises in particular solutions and exhibits
a possible divergence in the density of the fluid.
Apart from these singular solutions, we also find regular
ones that lead to avoidance of finite-distance singularities but only for the case of curved
branes. In particular, we may avoid finite-distance
singularities for a curved brane when $\gamma > -1/2$. In contrast with the previous
results, we do not find any range of $\gamma$ that leads to avoidance of finite-distance singularities
for a flat brane.
As we mentioned before, the existence of a general regular solution for a flat brane
implies that a self-tuning mechanism may be constructed. The failure of our flat-brane
model to offer such possibility leads us to its generalization which is implemented by
considering an interaction between the two components in the bulk. In this more complicated
case we find that for an adequate choice of the interaction parameters and for
$-1<\gamma<-1/2$, the avoidance of singularities is recovered.

Our approach in avoiding finite-distance singularities is to find ranges of the parameter
$\gamma$ and later on of the interaction coupling-coefficients that allow for the existence of
solutions that are singular only at infinite distance. We do not, however, consider
the possibility of constructing regular solutions with a matching mechanism as in
\cite{Forste2}. This mechanism is implemented by exploiting solutions that exhibit
a finite-distance singularity that is located only in the half line of the extra dimension
away from the position of the space of matching.
%

The structure of this paper is the following: In Section 2, we start
by giving a set up of the basic equations of our model consisting of
a brane in a bulk with a scalar field and an analog of perfect fluid.
These two bulk components may exchange energy in a way that the total energy is conserved. The field equations are written as a dynamical system which we analyze with the method of asymptotic splittings, cf. \cite{skot}. As a first step in Section 3, we focus on the case in which there is no exchange of energy between the bulk components and derive all possible asymptotic decompositions of the dynamical
system together with their dominant balances, i.e., the different
possible asymptotic modes of behavior. In particular Subsections 3.4-3.6, are devoted to the asymptotic structure of our braneworlds near finite-distance
singularities in the bulk, while in Subsection 3.7, we focus on behavior at infinity.
As a second step in Section 4 we consider that the two bulk components interact with each
other and resolve the unwanted situation discussed in Subsection 3.7.
We conclude and discuss our results in Section 5.
\section{Field equations}
We consider a braneworld model consisting of a three-brane embedded
in a five-dimensional bulk space that is filled with a massless
scalar field and an analog of perfect fluid. We assume a bulk metric of the
form \be \label{warpmetric} g_{5}=a^{2}(Y)g_{4}+dY^{2}, \ee where
$g_{4}$ is the four-dimensional flat, de Sitter or anti de Sitter
metric, i.e.,
\be \label{branemetrics}
g_{4}=-dt^{2}+f^{2}_{\kappa}g_{3}, \ee where \be
g_{3}=dr^{2}+h^{2}_{\kappa}g_{2} \ee and \be
g_{2}=d\theta^{2}+\sin^{2}(\theta)d\varphi^{2}.
\ee
Here $f_{\kappa}=1,\cosh (H t)/H,\cos (H t)/H$
($H^{-1}$ is the de Sitter curvature radius) and $h_{\kappa}=r,\sin r,\sinh r$,
respectively.
For the scalar field we assume an energy-momentum tensor of
the form $T^{1}_{AB}=(\rho_{1}+P_{1})u_{A}u_{B}-P_{1}g_{AB}$
where $A,B=1,2,3,4,5$, $u_{A}=(0,0,0,0,1)$ and $\rho_{1}$, $P_{1}$
is the density and pressure of the scalar field which we take as
$P_{1}=\rho_{1}=\lambda\phi'^{2}/2$, where the prime
denotes differentiation with respect to $Y$ and $\lambda$ is a
parameter. Respectively, the energy-momentum tensor of the fluid
is $T^{2}_{AB}=(\rho_{2}+P_{2})u_{A}u_{B}-P_{2}g_{AB}$ and we assume an
equation of state of the form $P_{2}=\gamma\rho_{2}$ between the
pressure $P_{2}$ and the density $\rho_{2}$ with $\gamma$ being a parameter.
All quantities $\rho_{1}$, $\rho_{2}$ and $P_{1}$, $P_{2}$ are functions of the fifth
dimension $Y$ only. The five-dimensional Einstein field equations, \be
G_{AB}=\kappa^{2}_{5}T_{AB}, \ee where $\kappa^{2}_{5}=M_{5}^{-3}$
and $M_{5}$ is the five dimensional Planck mass, can then be written
as
\bq
\label{orig_system1}
\dfrac{a''}{a}&=&-A\lambda\phi'^{2}-\dfrac{2}{3}A(1+2\gamma)\rho_{2},\\
\label{orig_system2}
\dfrac{a'^{2}}{a^{2}}&=&\dfrac{\lambda
A}{3}\phi'^{2}+ \dfrac{2A}{3} \rho_{2}+\dfrac{kH^{2}}{a^{2}}, \eq where
$A=\kappa_{5}^{2}/4$, $k=0,\pm 1$ (and the prime ($'$) denotes
differentiation with respect to $Y$).
We assume that there is an exchange of energy between the two matter components depending on the values and signs of the two constants $\nu,\sigma$, such that the total energy is conserved \cite{barclift}, so that we have the following two
equations,
\bq \label{orig_system3}
\lambda\phi'\phi''+4\lambda\dfrac{a'}{a}{\phi'}^{2}&=&-\frac{\lambda\nu}{2}
\frac{a'}{a}{\phi'}^{2}+\sigma\rho_{2}\frac{a'}{a},\\
\label{orig_system4}
\rho_{2}'+4(\gamma+1)\dfrac{a'}{a}\rho_{2}&=&\frac{\lambda\nu}{2}
\frac{a'}{a}{\phi'}^{2}-\sigma\rho_{2}\frac{a'}{a}.
\eq
Eqs. (\ref{orig_system1}) and (\ref{orig_system2}) are not independent, since
Eq. (\ref{orig_system1}) was derived after substitution of Eq. (\ref{orig_system2}) in
the field equation $G_{\a\a}=\kappa_{5}^{2}T_{\a\a}=4 A T_{\a\a}$, $\a=1,2,3,4$:
\be
\frac{a''}{a}+\frac{{a'}^{2}}{a^{2}}-\frac{k H^{2}}{a^{2}}=-\frac{2A}{3}\lambda{\phi'}^{2}-
\frac{4A}{3}\gamma\rho_{2}.
\ee
In our analysis we use the independent Eqs. (\ref{orig_system1}), (\ref{orig_system3}) and (\ref{orig_system4})
to determine the unknown variables $a$, $a'$, $\phi'$ and $\rho_{2}$,
while Eq. (\ref{orig_system2}) will play the role of a constraint equation for our system.

Our purpose is to find all possible asymptotic behaviors of
(general or particular) solutions of the system defined by the
dynamical equations (\ref{orig_system1})-(\ref{orig_system4}). The
most adequate tool for this quest is perhaps the method of
asymptotic splittings summarized in \cite{skot}. The first step is
to write this system in the form of a suitable dynamical system. We
introduce the following set of variables:
$$(x,y,z,w)=(a,a',\phi',\rho_{2}).$$
The system of equations
(\ref{orig_system1}), (\ref{orig_system3}) and (\ref{orig_system4})
then becomes the following dynamical system
\bq
\label{syst1_1_n_s}
x'&=&y\\
y'&=&-A \lambda z^{2}x-\dfrac{2}{3}A(1+2\gamma)wx\\
z'&=&-\left(4+\dfrac{\nu}{2}\right) \dfrac{y z}{x}+\frac{\sigma}{\lambda}\frac{y w}{x z}\\
\label{syst1_4_n_s}
w'&=&-(4(\gamma+1)+\sigma)\dfrac{y w}{x}+\frac{\lambda \nu}{2}\frac{y z^{2}}{x},
\eq
while equation (\ref{orig_system2}) now reads
\be
\label{constraint} \dfrac{y^{2}}{x^{2}}=\dfrac{A\lambda}{3}z^{2}+
\dfrac{2A}{3}w+\dfrac{kH^{2}}{x^{2}}.
\ee
Since this last equation does not contain derivatives with
respect to $Y$, it is a constraint equation for the system
(\ref{syst1_1_n_s})-(\ref{syst1_4_n_s}). The vector field defined
by the above system is given by
\be
\label{vectorfield_full_n_s}
\mathbf{f}= \left(y,-A \lambda z^{2}x-\dfrac{2}{3}A(1+2\gamma)w x,
-\left(4+\dfrac{\nu}{2}\right) \dfrac{y z}{x}+\frac{\sigma}{\lambda}\frac{y w}{x z},
-(4(\gamma+1)+\sigma)\dfrac{y w}{x}+\frac{\lambda \nu}{2}\frac{y z^{2}}{x}
\right)^{\top}.
\ee

Before we proceed with the analysis of the above system, we
introduce the following terminology for the possible singularities
to occur at a finite-distance from the brane. Specifically we call a
state where:
\begin{enumerate}
\item[i)] $a\rightarrow 0$, $a'\rightarrow \infty$, $\phi'\rightarrow \infty$,
$\rho_{2}\rightarrow 0, \rho_{s}, \infty$: a singularity of collapse
type I,
\item[ii)] $a\rightarrow 0$, $a'\rightarrow a'_{s}$, $\phi'\rightarrow 0$,
$\rho_{2}\rightarrow\rho_{s},\infty$: a singularity of collapse type II,
\item[iii)] $a\rightarrow \infty$, $a'\rightarrow -\infty$, $\phi'\rightarrow 0$,
$\rho_{2}\rightarrow\infty$: a big rip singularity,
\end{enumerate}
where $a'_{s}$ and $\rho_{s}$ 
are non-vanishing constants.

In the following Subsections we first analyze the case in which there is no
exchange of energy between the two components in the bulk, that is we take
$\nu=\sigma=0$, and later on we examine the very interesting case $\sigma =0$ and
$\nu$ arbitrary and we comment on the results of the case $\nu=0$ and $\sigma$ arbitrary.
The generic case $\sigma$, $\nu$ nonzero is difficult to study in generality and classify all
possible behaviors.
\section{Non-interacting mixture in the bulk}
In this Section we let $\nu=\sigma=0$ so that the system
Eqs. (\ref{syst1_1_n_s})-(\ref{syst1_4_n_s}) becomes
\bq
\label{syst1_1}
x'&=&y\\
y'&=&-A \lambda z^{2}x-\dfrac{2}{3}A(1+2\gamma)wx\\
z'&=&-4 \dfrac{y z}{x}\\
\label{syst1_4}
w'&=&-4(\gamma+1)\dfrac{y w}{x},
\eq
while equation (\ref{constraint}) remains the same. The vector field
of the above system is
\be
\label{vectorfield_full}
\mathbf{f}= \left(y,-A \lambda z^{2}x-\dfrac{2}{3}A(1+2\gamma)wx, -4
\dfrac{yz}{x},-4(\gamma+1)\dfrac{yw}{x}\right)^{\top},
\ee
and there are three possible ways of decomposing it. We analyze them in turn
in the following Subsections.
\subsection{Decomposition I}
The first way of decomposing the vector field
(\ref{vectorfield_full}) is to assume that its dominant part is
given by the form \be \label{dominantpart1}
\mathbf{f}^{(0)}=\left(y,-A \lambda z^{2}x, -4
\dfrac{yz}{x},-4(\gamma+1)\dfrac{yw}{x}\right)^{\top}, \ee while its candidate
subdominant part is: \be
\mathbf{f}^{(1)}=\left(0,-\dfrac{2}{3}A(1+2\gamma)wx,0,0\right)^{\top}.
\ee At this point we wish to determine all \emph{dominant balances},
that is pairs of the form \be
\mathcal{B}=\{\mathbf{a},\mathbf{p}\}, \quad \textrm{where} \quad
\mathbf{a}=(\alpha,\beta,c,\zeta), \quad \mathbf{p}=(p,q,r,s), \ee
with \be (p,q,r,s)\in\mathbb{Q}^{4} \quad \textrm{and} \quad
(\alpha,\beta,c,\zeta)\in
\mathbb{C}^{4}\smallsetminus\{\mathbf{0}\}, \ee that describe all
possible asymptotic behaviors around the assumed position of the
singularity at $Y_{s}$. We thus insert
\be
\label{variables}
(x,y,z,w)=(\alpha\Upsilon^{p},\beta
\Upsilon^{q},c\Upsilon^{r},\zeta \Upsilon^{s}),
\ee
where $\Upsilon=Y-Y_{s}$, into the asymptotic system defined by the first
decomposition, that is
\bq
\label{decopm1_1}
x'&=&y\\
\label{decopm1_2}
y'&=&-A \lambda z^{2}x\\
z'&=&-4 \dfrac{yz}{x}\\
\label{decopm1_4}
w'&=&-4(\gamma+1)\dfrac{yw}{x}.
\eq
This leads us to the list of all possible dominant balances. For each
balance we need to check that the \emph{dominance condition}, \be
\label{dominance1}
\lim_{\Upsilon\rightarrow 0}
\frac{\mathbf{f}^{(1)}(\mathbf{a}\Upsilon^{\mathbf{p}})}
{\Upsilon^{\mathbf{p}-1}}=0,
\ee
is satisfied and then discard those balances that do not satisfy Eq. (\ref{dominance1}). We
end up with the following acceptable balances\footnote{In the balance $_{I}\mathcal{B}_{1}$ the coefficient $\sqrt{3}/(4\sqrt{A\lambda})$ may also be $-\sqrt{3}/(4\sqrt{A\lambda})$.
This is true for every balance we find that has a square root in the coefficient of $\phi'$, but for
simplicity we examine these balances only for the $(+)$ sign.}:
\bq
\label{balances_I}
_{I}\mathcal{B}_{1}&=&\{(\alpha,\alpha/4,\sqrt{3}/(4\sqrt{A\lambda}),\zeta),
(1/4,-3/4,-1,-(\gamma+1))\},\quad \gamma < 1,\\
_{I}\mathcal{B}_{2}&=&\{(\alpha,\alpha/4,\sqrt{3}/(4\sqrt{A\lambda}),0),
(1/4,-3/4,-1,s)\},\\
_{I}\mathcal{B}_{3}&=&\{(\a,\a,0,\zeta),(1,0,-1,-4(\gamma+1))\},\quad \gamma\leq-1/2,\\
_{I}\mathcal{B}_{4}&=&\{(\a,\a,0,0),(1,0,-1,s)\},\\
_{I}\mathcal{B}_{5}&=&\{(\a,0,0,0),(0,-1,-1,s)\}.
\eq
The balance $_{I}\mathcal{B}_{5}$ leads to $\rho_{2}$ being identically zero which means that
it describes a behavior that applies in the case of bulk filled exclusively with the
scalar field. We studied this balance in our previous work in \cite{ack} and found it
to be unacceptable. We therefore do not consider it any further in this paper.

The balances $_{I}\mathcal{B}_{1-4}$ are exact solutions of the system
(\ref{decopm1_1})-(\ref{decopm1_4}). The constraint Eq. (\ref{constraint}) has contributed
in the Eq. (\ref{decopm1_2}) with all its terms excluding the term of the fluid density.
We can therefore substitute these balances in the constraint equation (\ref{constraint})
neglecting the term of the fluid density and find out if they correspond to a flat or curved brane. We find that for $\gamma\ne -1/2$,
the balances $_{I}\mathcal{B}_{1}$ and $_{I}\mathcal{B}_{2}$ correspond to a flat brane while
the balances $_{I}\mathcal{B}_{3}$ and $_{I}\mathcal{B}_{4}$ correspond to a
curved brane with the arbitrary constant $\alpha$ satisfying $\alpha^{2}=kH^{2}$.

The value $\gamma=-1/2$ is of special interest for our analysis since for this value of $\gamma$
the system (\ref{syst1_1})-(\ref{syst1_4}) identifies with the system of this first
decomposition (\ref{decopm1_1})-(\ref{decopm1_4}). This means that the balances
we find for $\gamma=-1/2$ are exact solutions of the system (\ref{syst1_1})-(\ref{syst1_4})
and they should therefore satisfy the entire constraint equation (\ref{constraint}).
These balances are: $_{I}\mathcal{B}_{2}$ (flat brane),  $_{I}\mathcal{B}_{4}$ (curved brane with
$\alpha^{2}=kH^{2}$), $_{I}\mathcal{B}_{3}$ (flat/curved brane with
$\zeta=3/(2A)(1-k H^{2}/\a^{2}$) and $_{I}\mathcal{B}_{1}$ (curved brane with $\zeta=-3/(2A)(kH^{2}/\a^{2})$).
\subsection{Decomposition II}
The second way of decomposing the vector field
(\ref{vectorfield_full}) is to take its dominant part to be
\be
\label{dominantpart2}
\mathbf{f}^{(0)}=\left(y,-\dfrac{2}{3}A(1+2\gamma)wx,-4
\dfrac{yz}{x}, -4(\gamma+1)\dfrac{yw}{x}\right)^{\top}.
\ee
Its candidate subdominant part reads,
\be
\mathbf{f}^{(1)}=(0,-A \lambda
z^{2}x,0,0)^{\top}.
\ee
For this second decomposition the system is
given by the following equations
\bq
\label{decopm2_1}
x'&=&y\\
\label{decopm2_2}
y'&=&-\dfrac{2}{3}A(1+2\gamma)wx\\
z'&=&-4 \dfrac{yz}{x}\\
\label{decopm2_4}
w'&=&-4(\gamma+1)\dfrac{yw}{x},
\eq
and the acceptable balances are calculated to be\footnote{The balance $_{II}\mathcal{B}_{4}$ is not analyzed any further
since it leads to $\phi'$ being identically zero and a similar argument applies as in
the case of $_{I}\mathcal{B}_{5}$ discussed above, in the decomposition I.}
\bq
\label{balances_II}
_{II}\mathcal{B}_{1}&=&\{(\alpha,\alpha p,c,3p^{2}/(2A)),
(p,p-1,-4p,-2)\}, \quad |\gamma|>1,\\
_{II}\mathcal{B}_{2}&=&\{(\alpha,\alpha p,0,3p^{2}/(2A)),
(p,p-1,r,-2)\}, \quad \gamma\neq -1,-1/2,\\
_{II}\mathcal{B}_{3}&=&\{(\a,\a,0,0),(1,0,r,-2)\}, \quad \gamma\neq -1/2,\\
_{II}\mathcal{B}_{4}&=&\{(\a,0,0,0),(0,-1,r,-2)\}, \quad \gamma\neq
-1/2,
\eq
where $p=1/(2(\gamma+1))$. Following the same trend as we did for Decomposition I,
we see that the constraint Eq. (\ref{constraint}) has contributed
in the Eq. (\ref{decopm2_2}) with all its terms excluding the term of the derivative of the
scalar field. We therefore substitute the balances $_{II}\mathcal{B}_{1-3}$ in the constraint equation
Eq. (\ref{constraint}) neglecting the term of the derivative of the scalar field and find that
the balances $_{II}\mathcal{B}_{1}$ and $_{II}\mathcal{B}_{2}$ correspond to a flat brane while
the balance $_{II}\mathcal{B}_{3}$ corresponds to a curved brane with $\alpha^{2}=kH^{2}$.
\subsection{Decomposition III}
The third way of decomposing the vector field
(\ref{vectorfield_full}) is to assume that all terms are dominant so
that
the system is given by Eqs. (\ref{syst1_1})-(\ref{syst1_4}). For
this third decomposition the dominant balances are \bq
\label{balances_III} _{III}\mathcal{B}_{1}&=&\{(\alpha,\alpha/4,c,
3/(32A)-\lambda c^{2}/2),
(1/4,-3/4,-1,-2)\}, \quad \gamma=1,\\
_{III}\mathcal{B}_{2}&=&\{(\alpha,\alpha p,0,3p^{2}/(2A)),
(p,p-1,-1,-2)\}, \quad \gamma\neq -1,-1/2,\\
_{III}\mathcal{B}_{3}&=&\{(\a,\a/4,\sqrt{3}/(4\sqrt{A\lambda}),0),(1/4,-3/4,-1,-2)\},
\\
_{III}\mathcal{B}_{4}&=&\{(\a,\a,0,0),(1,0,-1,-2)\},
\eq
with $p=1/(2(\gamma+1))$. These balances are
exact solutions of the system (\ref{syst1_1})-(\ref{syst1_4}) and
they should therefore satisfy the constraint equation
(\ref{constraint}). We find that the balances
$_{III}\mathcal{B}_{1}$, $_{III}\mathcal{B}_{2}$ and
$_{III}\mathcal{B}_{3}$ correspond to a flat brane,
while the balance $_{III}\mathcal{B}_{4}$ corresponds to a curved brane with
$\a^{2}=kH^{2}$. Notice that the three decompositions considered above
exhaust all possible asymptotic ways that the vector field (\ref{vectorfield_full}) can split.

Subsections 3.4-3.7 are the heart of this Section that focuses on non-interacting
bulk components. We have grouped the
possible balances $_{I-III}\mathcal{B}$ found above into four
different sets according to the type of singularity they lead to, or,
their regular behavior. For each particular balance we follow the
method of asymptotic splittings up to the point where we end up with
a well-defined series expansion. These expansions serve to
completely justify our claims that the asymptotic behavior of the
braneworld is the one claimed. We find that in all cases these
behaviors result from Puiseaux representations, in particular there
are no logarithmic terms present in any of the expansions.
\subsection{Collapse type I singularities}
In this Section, we analyze the balances $_{I}\mathcal{B}_{1}$,
$_{III}\mathcal{B}_{3}$, $_{I}\mathcal{B}_{2}$,
$_{II}\mathcal{B}_{2}$, $_{III}\mathcal{B}_{1}$ and
$_{III}\mathcal{B}_{2}$ that describe the asymptotics around
collapse type I singularities.
\subsubsection{The balance $_{I}\mathcal{B}_{1}$}
We start with the analysis of the balance $_{I}\mathcal{B}_{1}$ that corresponds to a flat
brane for $\gamma\neq -1/2$. We will show that for different values of $\gamma$ this
balance implies
different behaviors of the matter density of the fluid around a
collapse type I singularity. We first have to calculate for this
balance the $\mathcal{K}$-matrix given by \be
_{I}\mathcal{K}_{1}=D\mathbf{f}^{(0)}(\mathbf{a})-\textrm{diag}\,
\mathbf{p}, \ee where $D\mathbf{f}^{(0)}(\mathbf{a})$ is the
Jacobian matrix of the dominant part $\mathbf{f}^{(0)}$ in Eq.
(\ref{dominantpart1}), \be D\mathbf{f}^{(0)}(x,y,z,w)=\left(
                     \begin{array}{cccc}
                       0        & 1                       & 0              & 0 \\ \\
               -\lambda A z^{2} & 0                       & -2\lambda A z x & 0 \\ \\
            4\dfrac{yz}{x^{2}} & -4\dfrac{z}{x}          & -4\dfrac{y}{x} & 0 \\ \\
  4(\gamma +1)\dfrac{yw}{x^{2}} & -4(1+\gamma)\dfrac{w}{x}& 0              &-4(1+\gamma)\dfrac{y}{x}\\
                     \end{array}
                   \right),
\ee evaluated on $\mathbf{a}$. We have that
$\mathbf{a}=(\alpha,\alpha/4,\sqrt{3}/(4\sqrt{A\lambda}),\zeta)$ and
$\mathbf{p}=(1/4,-3/4,-1,-(\gamma+1))$, so that the
$\mathcal{K}$-matrix in this case is
\beq
_{I}\mathcal{K}_{1}&=&D\mathbf{f}^{(0)}
((\alpha,\alpha/4,\sqrt{3}/(4\sqrt{A\lambda}),\zeta))
-\textrm{diag}(1/4,-3/4,-1,-(\gamma+1))=\\ \\
&=&\left(
  \begin{array}{cccc}
    -\dfrac{1}{4} & 1 & 0 & 0\\ \\
    -\dfrac{3}{16} & \dfrac{3}{4} & -\dfrac{\a\sqrt{3A\lambda}}{2} & 0\\ \\
    \dfrac{\sqrt{3}}{4\a\sqrt{A\lambda}} & -\dfrac{\sqrt{3}}{\a\sqrt{A\lambda}} & 0 & 0\\ \\
    (1+\gamma)\dfrac{\zeta}{\a} & -4(1+\gamma)\dfrac{\zeta}{\a} & 0 & 0\\
  \end{array}
\right).\\ \eeq Next, we calculate the $\mathcal{K}$-exponents for
this balance. These exponents are the eigenvalues of the matrix
$_{I}\mathcal{K}_{1}$ and constitute its spectrum,
$spec(_{I}\mathcal{K}_{1})$. We wish to build series expansions of
the variables in the form \be \label{Puiseux}
\mathbf{x}=\Upsilon^{\mathbf{p}}(\mathbf{a}+
\Sigma_{j=1}^{\infty}\mathbf{c}_{j}\Upsilon^{j/S}), \ee where
$\mathbf{x}=(x,y,z,w)$,
$\mathbf{c}_{j}=(c_{j1},c_{j2},c_{j3},c_{j4})$, and $S$ is the
least common multiple of the denominators of the positive
$\mathcal{K}$-exponents and the non-dominant exponents $q^{(1)}$
defined by the requirement \be \label{sub_exp}
\frac{\mathbf{f}^{(1)}(\Upsilon^{\mathbf{p}})}
{\Upsilon^{\mathbf{p}-1}}\sim\Upsilon ^{q^{(1)}}, \ee
(cf. \cite{skot}, \cite{goriely}). 
The arbitrary constants of any particular or general solution
first appear in those terms in the series (\ref{Puiseux}) whose
coefficients $\mathbf{c}_{k}$ have indices $k=\varrho S$, where
$\varrho$ is a non-negative $\mathcal{K}$-exponent. The number of
non-negative $\mathcal{K}$-exponents therefore equals the number
of arbitrary constants that appear in the series expansions of
(\ref{Puiseux}). There is always the $-1$ exponent that
corresponds to an arbitrary constant that is the position of the
singularity, $Y_{s}$.

The balance $_{I}\mathcal{B}_{1}$
corresponds thus to a general solution in our case if and only if
it possesses three non-negative $\mathcal{K}$-exponents (the
fourth arbitrary constant is the position of the singularity,
$Y_{s}$). Actually, once we use the constraint equation (\ref{constraint}), one of the
three arbitrary constants corresponding to the three non-negative
$\mathcal{K}$-exponents will be set to a specific value, so that the general solution of the
dynamical system (\ref{syst1_1})-(\ref{syst1_4}) with constraint equation (\ref{constraint})
will exhibit three in total arbitrary constants taken into account also the singularity position
$Y_{s}$. Here we find
\be
\textrm{spec}(_{I}\mathcal{K}_{1})=\{-1,0,0,3/2\}.
\ee
The double multiplicity of the zero $\mathcal{K}$-exponent reflects the fact
that there are two arbitrary constants in this dominant balance.
In total we have three non-negative $\mathcal{K}$-exponents which
means that this balance indeed corresponds to a general solution.

Naturally, the behavior of $\rho_{2}$ depends on the exponent
$-(\gamma+1)$. We try inserting different values of $\gamma$ so that
to trace all possible asymptotics for $\rho_{2}$ keeping in mind that
$\gamma \leq 1$. For instance, for $\gamma=0$ we have that \be
_{I}\mathcal{B}_{1}=\{(\alpha,\alpha/4,\sqrt{3}/(4\sqrt{A\lambda}),\zeta),
(1/4,-3/4,-1,-1)\} \ee and substituting in the system
(\ref{syst1_1})-(\ref{syst1_4}) the particular value $\gamma=0$ and
the forms \be \nonumber
x=\Sigma_{j=0}^{\infty}c_{j1}\Upsilon^{j/2+1/4}, \quad
y=\Sigma_{j=0}^{\infty}c_{j2}\Upsilon^{j/2-3/4}, \quad
z=\Sigma_{j=0}^{\infty}c_{j3}\Upsilon^{j/2-1},   \quad
w=\Sigma_{j=0}^{\infty}c_{j4}\Upsilon^{j/2-1}, \ee
we arrive at the following asymptotic expansions: \bq
\label{puiseux_I_1_0_i}
x&=&\a\Upsilon ^{1/4}+\a\zeta/3\Upsilon^{5/4}+c_{31}\Upsilon^{7/4}+\ldots,\\
y&=&\a/4\Upsilon ^{-3/4}+5\a\zeta/12\Upsilon^{1/4}+7/4 c_{31}\Upsilon^{3/4}+\ldots,\\
\label{puiseux_I_1_0_iii}
z&=&\sqrt{3}/(4\sqrt{A\lambda})\Upsilon ^{-1}-\zeta\sqrt{3}/(3\sqrt{A\lambda})-c_{31}\sqrt{3}/(\a \sqrt{A\lambda})\Upsilon^{1/2}+\ldots,\\
\label{puisuex_I_1_0_iv}
w&=&\zeta\Upsilon^{-1}-4\zeta^{2}/3-4\zeta c_{31}/\a\Upsilon^{1/2}+\ldots
\, . \eq
The next step in our analysis is to check if for each $j$
satisfying $j/2=\varrho$ with $\varrho$ a positive eigenvalue, the
corresponding eigenvector $v$ of the transpose of the $_{I}\mathcal{K}_{1}$
matrix is such that the compatibility conditions hold, namely,
\be
v^{\top}\cdot P_{j}=0,
\ee
where $P_{j}$ are polynomials in $\mathbf{c}_{1},\ldots, \mathbf{c}_{j-1}$
given by
\be
(_{I}\mathcal{K}_{1}-(j/2)\mathcal{I})\mathbf{c}_{j}=P_{j}.
\ee
Here the corresponding relation $j/2=3/2$, is valid only for $j=3$ and
the compatibility condition indeed holds since, \be
(_{I}\mathcal{K}_{1}-3/2\mathcal{I})\mathbf{c}_{3}= \left(
  \begin{array}{cccc}
    -\dfrac{7}{4} & 1 & 0 & 0\\ \\
    -\dfrac{3}{16} & -\dfrac{3}{4} & -\dfrac{\a\sqrt{3A\lambda}}{2} & 0\\ \\
    \dfrac{\sqrt{3}}{4\a\sqrt{A\lambda}} & -\dfrac{\sqrt{3}}{\a\sqrt{A\lambda}} & -\dfrac{3}{2} & 0\\ \\
    \dfrac{\zeta}{\a} & -4\dfrac{\zeta}{\a} & 0 & -\dfrac{3}{2}\\
  \end{array}
\right) \left(
  \begin{array}{c}
    c_{31} \\ \\
    \dfrac{7}{4} c_{31} \\ \\
     -\dfrac{\sqrt{3}}{\a \sqrt{A\lambda}}c_{31}\\ \\
     -\dfrac{4\zeta}{\a}c_{31}\\
  \end{array}
\right)=\left(
          \begin{array}{c}
            0 \\ \\
            0 \\ \\
            0\\ \\
            0 \\
          \end{array}
        \right).
\ee This means that a representation of the solution asymptotically
with a Puiseux series as this is given by Eqs.
(\ref{puiseux_I_1_0_i})-(\ref{puisuex_I_1_0_iv}) is valid. We thus
conclude that as $Y\rightarrow Y_{s}$, or equivalently as
$\Upsilon\rightarrow 0$, \be a\rightarrow 0,\quad a'\rightarrow
\infty,\quad \phi'\rightarrow\infty, \quad \rho_{2}\rightarrow\infty.
\ee

We consider also the case $\gamma=-1/2$ for which we find
\bq
\label{puiseux_I_1_-1/2_i}
x&=&\a\Upsilon ^{1/4}-\a\sqrt{A\lambda}/\sqrt{3}c_{33}\Upsilon^{7/4}+\ldots,\\
y&=&\a/4\Upsilon ^{-3/4}-7\a\sqrt{A\lambda}/(4\sqrt{3})c_{33}\Upsilon^{3/4}+\ldots,\\
\label{puiseux_I_1_-1/2_iii}
z&=&\sqrt{3}/(4\sqrt{A\lambda})\Upsilon ^{-1}+c_{33}\Upsilon^{1/2}+\ldots,\\
\label{puisuex_I_1_-1/2_iv}
w&=&\zeta\Upsilon^{-1/2}+2\zeta\sqrt{A\lambda}/\sqrt{3}c_{33}\Upsilon+\ldots
\, ,
\eq
where $\zeta=-3k H^{2}/(2A\a^{2})$. The compatibility condition for $j=3$ is satisfied
in the following way:
\be
(_{I}\mathcal{K}_{1}-3/2\mathcal{I})\mathbf{c}_{3}= \left(
  \begin{array}{cccc}
    -\dfrac{7}{4} & 1 & 0 & 0\\ \\
    -\dfrac{3}{16} & -\dfrac{3}{4} & -\dfrac{\a\sqrt{3A\lambda}}{2} & 0\\ \\
    \dfrac{\sqrt{3}}{4\a\sqrt{A\lambda}} & -\dfrac{\sqrt{3}}{\a\sqrt{A\lambda}} & -\dfrac{3}{2} & 0\\ \\
    \dfrac{\zeta}{2\a} & -2\dfrac{\zeta}{\a} & 0 & -\dfrac{3}{2}\\
  \end{array}
\right) \left(
  \begin{array}{c}
    -\dfrac{\sqrt{A\lambda}\a}{\sqrt{3}}c_{33} \\ \\
    -\dfrac{7}{4}\dfrac{\sqrt{A\lambda}\a}{\sqrt{3}}c_{33} \\ \\
     c_{33}\\ \\
     2\zeta \dfrac{\sqrt{A\lambda}}{\sqrt{3}}c_{33}\\
  \end{array}
\right)=\left(
          \begin{array}{c}
            0 \\ \\
            0 \\ \\
            0\\ \\
            0 \\
          \end{array}
        \right).
\ee
We therefore see that for $\gamma=-1/2$
Eqs (\ref{puiseux_I_1_-1/2_i})-(\ref{puisuex_I_1_-1/2_iv}) express
the asymptotic behavior that corresponds to a curved brane around a collapse I singularity
with the density of the fluid diverging, i.e as
$\Upsilon\rightarrow 0$,
\be a\rightarrow 0,\quad a'\rightarrow \infty,\quad
\phi'\rightarrow\infty, \quad \rho_{2}\rightarrow\infty.
\ee

For $\gamma=-2$ we find a different behavior
\bq
\label{puiseux_I_1_-2_i}
x&=&\a\Upsilon ^{1/4}-\sqrt{A\lambda}\a/\sqrt{3}c_{33}\Upsilon^{7/4}+\ldots,\\
y&=&\a/4\Upsilon ^{-3/4}-7\sqrt{A\lambda}\a/(4\sqrt{3})c_{33}\Upsilon^{3/4}+\ldots,\\
z&=&\sqrt{3}/(4\sqrt{A\lambda})\Upsilon ^{-1}+c_{33}\Upsilon^{1/2}+\ldots,\\
\label{puiseux_I_1_-2_iv}
w&=&\zeta\Upsilon-4\zeta\sqrt{A\lambda}/\sqrt{3}c_{33}\Upsilon^{5/2}+\ldots.
\eq We ought to check the compatibility condition for $j=3$. We find
that this is trivially satisfied since \be
(_{I}\mathcal{K}_{1}-3/2\mathcal{I})\mathbf{c}_{3}= \left(
  \begin{array}{cccc}
    -\dfrac{7}{4}  & 1 & 0 & 0\\ \\
    -\dfrac{3}{16} & -\dfrac{3}{4} & -\dfrac{\a\sqrt{3A\lambda}}{2} & 0\\ \\
    \dfrac{\sqrt{3}}{4\a\sqrt{A\lambda}} & -\dfrac{\sqrt{3}}{\a\sqrt{A\lambda}} & -\dfrac{3}{2} & 0\\ \\
    -\dfrac{\zeta}{\a} & 4\dfrac{\zeta}{\a} & 0 & -\dfrac{3}{2}\\
  \end{array}
\right) \left(
  \begin{array}{c}
    -\dfrac{\sqrt{A\lambda}\a}{\sqrt{3}}c_{33} \\ \\
    -\dfrac{7}{4}\dfrac{\sqrt{A\lambda}\a}{\sqrt{3}}c_{33} \\ \\
     c_{33}\\ \\
     -4\zeta \dfrac{\sqrt{A\lambda}}{\sqrt{3}}c_{33}\\
  \end{array}
\right)=\left(
          \begin{array}{c}
            0 \\ \\
            0 \\ \\
            0\\ \\
            0 \\
          \end{array}
        \right).
\ee
The relations (\ref{puiseux_I_1_-2_i})-(\ref{puiseux_I_1_-2_iv})
are therefore valid representations of a general solution around the
singularity at $Y_{s}$. We can therefore conclude that as
$\Upsilon\rightarrow 0$, \be a\rightarrow 0, \quad a'\rightarrow
\infty, \quad \phi'\rightarrow\infty,\quad \rho_{2}\rightarrow 0. \ee

A yet different behavior is met for $\gamma=-1$: \bq
\label{puiseux_I_1_-1_i}
x&=&\a\Upsilon ^{1/4}-\sqrt{A\lambda}\a/\sqrt{3}c_{33}\Upsilon^{7/4}+\ldots,\\
y&=&\a/4\Upsilon ^{-3/4}-7\sqrt{A\lambda}\a/(4\sqrt{3})c_{33}\Upsilon^{3/4}+\ldots,\\
z&=&\sqrt{3}/(4\sqrt{A\lambda})\Upsilon ^{-1}+c_{33}\Upsilon^{1/2}+\ldots,\\
\label{puiseux_I_1_-1_iv} w&=&\zeta+\ldots. \eq The compatibility
condition at $j=3$ holds true since we find \be
(_{I}\mathcal{K}_{1}-3/2\mathcal{I})\mathbf{c}_{3}= \left(
  \begin{array}{cccc}
    -\dfrac{7}{4} & 1 & 0 & 0\\ \\
    -\dfrac{3}{16} & -\dfrac{3}{4} & -\dfrac{\a\sqrt{3A\lambda}}{2} & 0\\ \\
    \dfrac{\sqrt{3}}{4\a\sqrt{A\lambda}} & -\dfrac{\sqrt{3}}{\a\sqrt{A\lambda}} & -\dfrac{3}{2} & 0\\ \\
      0 & 0 & 0 & -\dfrac{3}{2}\\
  \end{array}
\right) \left(
  \begin{array}{c}
    -\dfrac{\sqrt{A\lambda}\a}{\sqrt{3}}c_{33} \\ \\
    -\dfrac{7}{4}\dfrac{\sqrt{A\lambda}\a}{\sqrt{3}}c_{33} \\ \\
     c_{33}\\ \\
     0\\
  \end{array}
\right)=\left(
          \begin{array}{c}
            0 \\ \\
            0 \\ \\
            0\\ \\
            0 \\
          \end{array}
        \right),
\ee so that the relations
(\ref{puiseux_I_1_-1_i})-(\ref{puiseux_I_1_-1_iv}) affirm that as
$\Upsilon\rightarrow 0$, \be a\rightarrow 0, \quad a'\rightarrow
\infty,\quad \phi'\rightarrow\infty, \quad \rho_{2}\rightarrow\zeta. \ee
\subsubsection{The balance $_{III}\mathcal{B}_{3}$}
We now move on to examine the next balance $_{III}\mathcal{B}_{3}=
\{(\a,\a/4,\sqrt{3}/(4\sqrt{A\lambda}),0), (1/4,-3/4,-1,-2)\}$ which
corresponds to a potentially general solution of a flat brane.
The $\mathcal{K}$-matrix for this balance is
\be
_{III}\mathcal{K}_{3}=D\mathbf{f}(\a,\a/4,\sqrt{3}/(4\sqrt{A\lambda}),0)
-\textrm{diag}(1/4,-3/4,-1,-2),
\ee
where $D\mathbf{f}$ is the Jacobian matrix of the vector field $\mathbf{f}$ in
Eq. (\ref{vectorfield_full}). The eigenvalues of the $_{III}\mathcal{K}_{3}$ matrix are
\be
\textrm{spec}(_{III}\mathcal{K}_{3})=\{-1,0,3/2,1-\gamma\}.
\ee
Setting $\gamma=0$ we get 
\be
\textrm{spec}(_{III}\mathcal{K}_{3})=\{-1,0,3/2,1\}.
\ee
We therefore have three non-negative $\mathcal{K}$-exponents which
means that in this case the balance indeed corresponds to a general
solution. The asymptotic expansions of the variables in this case
are
\bq
\label{puisuex_1_2_0_i}
x&=&\a\Upsilon^{1/4}+2/3 A\a c_{24}\Upsilon^{5/4}-\a\sqrt{A\lambda/3}c_{33}\Upsilon^{7/4}+\ldots,\\
y&=&\a/4\Upsilon^{-3/4}+5/6A\a c_{24}\Upsilon^{1/4}-7/4\a\sqrt{A\lambda/3}c_{33}\Upsilon^{3/4}+\ldots.\\
z&=&\sqrt{3}/(4\sqrt{A\lambda})\Upsilon^{-1}-2\sqrt{A/(3\lambda)}c_{24}+c_{33}\Upsilon^{1/2}+\ldots,\\
\label{puiseux_1_2_0_iv}
w&=&c_{24}\Upsilon^{-1}+\ldots. \eq We have to check the compatibility conditions
for $j=2$ and $j=3$. Since we
find \be (_{III}\mathcal{K}_{3}-\mathcal{I})\mathbf{c}_{2}= \left(
  \begin{array}{cccc}
    -\dfrac{5}{4} & 1 & 0 & 0\\ \\
    -\dfrac{3}{16} & -\dfrac{1}{4} & -\dfrac{\a\sqrt{3A\lambda}}{2} & -(2/3)A \a\\ \\
    \dfrac{\sqrt{3}}{4\a\sqrt{A\lambda}} & -\dfrac{\sqrt{3}}{\a\sqrt{A\lambda}} & -1 & 0\\ \\
    0 & 0 & 0 & 0\\
  \end{array}
\right) \left(
  \begin{array}{c}
    \dfrac{2A\a}{3}c_{24} \\ \\
    \dfrac{5A\a}{6}c_{24} \\ \\
     -\dfrac{2\sqrt{A}}{\sqrt{3\lambda}}c_{24}\\ \\
     c_{24}\\
  \end{array}
\right)=\left(
          \begin{array}{c}
            0 \\ \\
            0 \\ \\
            0\\ \\
            0 \\
          \end{array}
        \right)
\ee
and
\be
(_{III}\mathcal{K}_{3}-3/2\mathcal{I})\mathbf{c}_{3}=
\left(
  \begin{array}{cccc}
    -\dfrac{7}{4} & 1 & 0 & 0\\ \\
    -\dfrac{3}{16} & -\dfrac{3}{4} & -\dfrac{\a\sqrt{3A\lambda}}{2} & -(2/3)A \a\\ \\
    \dfrac{\sqrt{3}}{4\a\sqrt{A\lambda}} & -\dfrac{\sqrt{3}}{\a\sqrt{A\lambda}} & -\dfrac{3}{2} & 0\\ \\
    0 & 0 & 0 & -\dfrac{1}{2}\\
  \end{array}
\right) \left(
  \begin{array}{c}
    -\dfrac{\sqrt{A\lambda}\a}{\sqrt{3}}c_{33} \\ \\
    -\dfrac{7}{4}\dfrac{\sqrt{A\lambda}\a}{\sqrt{3}}c_{33} \\ \\
     c_{33}\\ \\
     0\\
  \end{array}
\right)=\left(
          \begin{array}{c}
            0 \\ \\
            0 \\ \\
            0\\ \\
            0 \\
          \end{array}
        \right),
\ee
the compatibility conditions for $j=2$ and $j=3$ hold true. Eqs.
(\ref{puisuex_1_2_0_i})-(\ref{puiseux_1_2_0_iv}) then imply that as
$\Upsilon\rightarrow 0$,
\be
a\rightarrow 0, \quad a'\rightarrow
\infty,\quad \phi'\rightarrow\infty, \quad \rho_{2}\rightarrow\infty.
\ee
In order to be able to compare and contrast our results found in
this Subsection with the ones that are presented later in Subsection 3.7,
we find it necessary to analyze here two more values of $\gamma$, namely $\gamma=-1/2$
and $\gamma=-3/4$. For $\gamma=-1/2$ we find that the
eigenvalues of the $_{III}\mathcal{K}_{3}$ matrix read
\be
\textrm{spec}(_{III}\mathcal{K}_{3})=\{-1,0,3/2,3/2\}.
\ee
For this value of $\gamma$ we find the asymptotic behavior
\bq \label{puisuex_3_3_-1/2_i}
x&=&\a\Upsilon^{1/4}-\sqrt{A \lambda}/\sqrt{3} \a c_{33}\Upsilon^{7/4}+\ldots,\\
y&=&\a/4\Upsilon^{-3/4}-7\sqrt{A\lambda}/(4\sqrt{3})\a c_{33}\Upsilon^{3/4}+\ldots,\\
z&=&\sqrt{3}/(4\sqrt{A\lambda})\Upsilon^{-1}+c_{33}\Upsilon^{1/2}+\ldots,\\
\label{puiseux_3_3_-1/2_iv}
w&=&c_{34}\Upsilon^{-1/2}+\ldots .
\eq
Since
\be
(_{III}\mathcal{K}_{3}-3/2\mathcal{I})\mathbf{c}_{3}=
\left(
  \begin{array}{cccc}
    -\dfrac{7}{4} & 1 & 0 & 0\\ \\
    -\dfrac{3}{16} & -\dfrac{3}{4} & -\dfrac{\a\sqrt{3A\lambda}}{2} & 0\\ \\
    \dfrac{\sqrt{3}}{4\a\sqrt{A\lambda}} & -\dfrac{\sqrt{3}}{\a\sqrt{A\lambda}} & -\dfrac{3}{2} & 0\\ \\
    0 & 0 & 0 & 0\\
  \end{array}
\right) \left(
  \begin{array}{c}
    -\dfrac{\sqrt{A\lambda}\a}{\sqrt{3}}c_{33} \\ \\
    -\dfrac{7\sqrt{A\lambda}\a}{4\sqrt{3}}c_{33} \\ \\
     c_{33}\\ \\
     c_{34}\\
  \end{array}
\right)=\left(
          \begin{array}{c}
            0 \\ \\
            0 \\ \\
            0\\ \\
            0 \\
          \end{array}
        \right),
\ee
the compatibility condition for $j=3$ is trivially satisfied, and
as it follows from Eqs. (\ref{puisuex_3_3_-1/2_i})-(\ref{puiseux_3_3_-1/2_iv})
as $\Upsilon\rightarrow 0$,
\be
a\rightarrow 0, \quad a'\rightarrow
\infty,\quad \phi'\rightarrow\infty, \quad \rho_{2}\rightarrow\infty.
\ee
Finally, for $\gamma=-3/4$ the eigenvalues become
\be
\textrm{spec}(_{III}\mathcal{K}_{3})=\{-1,0,3/2,7/4\}, \ee while
the asymptotic behavior now is
\bq
\label{puisuex_3_3_-3/4_i}
x&=&\a\Upsilon^{1/4}-\sqrt{A\lambda}/\sqrt{3}\a
c_{63}\Upsilon^{7/4}+16/33 A\a c_{74}\Upsilon^{2}+\ldots,\\
y&=&\a/4\Upsilon^{-3/4}-7\sqrt{A\lambda}/(4\sqrt{3})\a
c_{63}\Upsilon^{3/4}+32/33 A\a c_{74}\Upsilon+\ldots\\
z&=&\sqrt{3}/(4\sqrt{A\lambda})\Upsilon^{-1}+c_{63}\Upsilon^{1/2}-
16\sqrt{3A}/(33\sqrt{\lambda})c_{74}\Upsilon^{3/4}+\ldots,\\
\label{puiseux_3_3_-3/4_iv} w&=&c_{74}\Upsilon^{-1/4}+\ldots .
\eq
We check the compatibility conditions for $j=6$ (note that here
$S=4$ and we see that for the eigenvalue $\varrho=3/2$ the
corresponding arbitrary constant appears when $j=\varrho S=6$) and
for $j=7$ are again trivially satisfied since we find that:
\be
(_{III}\mathcal{K}_{3}-3/2\mathcal{I})\mathbf{c}_{6}=
\left(
  \begin{array}{cccc}
    -\dfrac{7}{4} & 1 & 0 & 0\\ \\
    -\dfrac{3}{16} & -\dfrac{3}{4} & -\dfrac{\a\sqrt{3A\lambda}}{2} & 1/3 A\a\\ \\
    \dfrac{\sqrt{3}}{4\a\sqrt{A\lambda}} & -\dfrac{\sqrt{3}}{\a\sqrt{A\lambda}} & -\dfrac{3}{2} & 0\\ \\
    0 & 0 & 0 & 1/4\\
  \end{array}
\right) \left(
  \begin{array}{c}
    -\dfrac{\sqrt{A\lambda}}{\sqrt{3}}\a c_{63} \\
    -\dfrac{7\sqrt{A\lambda}}{4\sqrt{3}}\a c_{63} \\ \\
     c_{63}\\ \\
     0\\
  \end{array}
\right)=\left(
          \begin{array}{c}
            0 \\ \\
            0 \\ \\
            0\\ \\
            0 \\
          \end{array}
        \right)
\ee
and
\be
(_{III}\mathcal{K}_{3}-7/4\mathcal{I})\mathbf{c}_{7}=
\left(
  \begin{array}{cccc}
    -2 & 1 & 0 & 0\\ \\
    -\dfrac{3}{16} & -1 & -\dfrac{\a\sqrt{3A\lambda}}{2} & 1/3 A\a\\ \\
    \dfrac{\sqrt{3}}{4\a\sqrt{A\lambda}} & -\dfrac{\sqrt{3}}{\a\sqrt{A\lambda}} & -\dfrac{7}{4} & 0\\ \\
    0 & 0 & 0 & 0\\
  \end{array}
\right)
 \left(
  \begin{array}{c}
    \dfrac{16}{33}A\a c_{74}\\ \\
    \dfrac{32}{33}A\a c_{74} \\ \\
    -\dfrac{16}{33}\sqrt{3A/\lambda}c_{74}\\ \\
     c_{74}\\
  \end{array}
\right)=
\left(\begin{array}{c}
            0 \\ \\
            0 \\ \\
            0 \\ \\
            0 \\
          \end{array}
        \right).
\ee
Hence Eqs. (\ref{puisuex_3_3_-3/4_i})-(\ref{puiseux_3_3_-3/4_iv})
show that as $\Upsilon\rightarrow 0$,
\be
a\rightarrow 0, \quad a'\rightarrow
\infty,\quad \phi'\rightarrow\infty, \quad \rho_{2}\rightarrow\infty.
\ee
\subsubsection{The balance $_{I}\mathcal{B}_{2}$}
Here we consider the balance $_{I}\mathcal{B}_{2}$ that corresponds to a flat brane.
The eigenvalues of the $\mathcal{K}$-matrix for this balance are
\be
\textrm{spec}(_{I}\mathcal{K}_{2})=\{-1,0,3/2,-1-\gamma-s\}.
\ee
For $s=1/2$ and $\gamma=-5/2$ we get
\be
\textrm{spec}(_{I}\mathcal{K}_{2})=\{-1,0,3/2,1\}.
\ee
We then find the following expansions
\bq
\label{puiseux_1_2_0_1/2_i}
x&=&\a\Upsilon^{1/4}-\sqrt{A\lambda/3}\a c_{33}\Upsilon^{7/4}+\ldots\\
y&=&\a/4\Upsilon^{-3/4}-7/4\sqrt{A\lambda/3}\a c_{33}\Upsilon^{3/4}+\ldots\\
z&=&\sqrt{3}/(4 \sqrt{A\lambda})\Upsilon^{-1}+c_{33}\Upsilon^{1/2}+\ldots\\
\label{puiseux_1_2_0_1/2_iv} w&=&c_{24}\Upsilon^{3/2}+\ldots \eq We
have to check the compatibility conditions for $j=2$ and $j=3$.
Since \be (_{I}\mathcal{K}_{2}-\mathcal{I})\mathbf{c}_{2}= \left(
  \begin{array}{cccc}
    -\dfrac{5}{4} & 1 & 0 & 0\\ \\
    -\dfrac{3}{16} & -\dfrac{1}{4} & -\dfrac{\a\sqrt{3A\lambda}}{2} & 0\\ \\
    \dfrac{\sqrt{3}}{4\a\sqrt{A\lambda}} & -\dfrac{\sqrt{3}}{\a\sqrt{A\lambda}} & -1 & 0\\ \\
    0 & 0 & 0 & 0\\
  \end{array}
\right) \left(
  \begin{array}{c}
     0 \\ \\
     0 \\ \\
     0\\ \\
     c_{24}\\
  \end{array}
\right)=\left(
          \begin{array}{c}
            0 \\ \\
            0 \\ \\
            0\\ \\
            0 \\
          \end{array}
        \right)
\ee and \be (_{I}\mathcal{K}_{2}-3/2\mathcal{I})\mathbf{c}_{3}=
\left(
  \begin{array}{cccc}
    -\dfrac{7}{4} & 1 & 0 & 0\\ \\
    -\dfrac{3}{16} & -\dfrac{3}{4} & -\dfrac{\a\sqrt{3A\lambda}}{2} & 0\\ \\
    \dfrac{\sqrt{3}}{4\a\sqrt{A\lambda}} & -\dfrac{\sqrt{3}}{\a\sqrt{A\lambda}} & -\dfrac{3}{2} & 0\\ \\
    0 & 0 & 0 & -\dfrac{1}{2}\\
  \end{array}
\right) \left(
  \begin{array}{c}
    -\dfrac{\sqrt{A\lambda}\a}{\sqrt{3}}c_{33} \\ \\
    -\dfrac{7}{4}\dfrac{\sqrt{A\lambda}\a}{\sqrt{3}}c_{33} \\ \\
     c_{33}\\ \\
     0\\
  \end{array}
\right)=\left(
          \begin{array}{c}
            0 \\ \\
            0 \\ \\
            0\\ \\
            0 \\
          \end{array}
        \right),
\ee the compatibility conditions are indeed satisfied and from Eqs.
(\ref{puiseux_1_2_0_1/2_i})-(\ref{puiseux_1_2_0_1/2_iv}) we see that
as $\Upsilon\rightarrow 0$, \be a\rightarrow 0, \quad a'\rightarrow
\infty,\quad \phi'\rightarrow\infty, \quad \rho_{2}\rightarrow 0. \ee
\subsubsection{The balance $_{III}\mathcal{B}_{1}$}
The balance $_{III}\mathcal{B}_{1}$ that corresponds to a
potentially general solution of a flat brane is valid only for
$\gamma=1$ and describes the asymptotics around a collapse type I
singularity. Here the eigenvalues of the matrix $_{III}\mathcal{K}_{1}$ are
\be
\textrm{spec}(_{III}\mathcal{K}_{1})=\{-1,0,0,3/2\}.
\ee
Since we find three non-negative $\mathcal{K}$-exponents for this balance we
conclude that it indeed corresponds to a general solution.

Here we find the following expressions \bq \label{puiseux_III_1_1_i}
x&=&\a\Upsilon^{1/4}-\a/(4c)c_{33}\Upsilon^{7/4}+\ldots\\
y&=&\a/4\Upsilon^{-3/4}-7\a/(16c)c_{33}\Upsilon^{3/4}+\ldots\\
z&=&c\Upsilon^{-1}+c_{33}\Upsilon^{1/2}+\ldots\\
\label{puiseux_III_1_1_iv} w&=&(3/(32A)-\lambda
c^{2}/2)\Upsilon^{-2}+ (2/c)(3/(32A)-\lambda c^{2}/2)
c_{33}\Upsilon^{-1/2}+\ldots\, \eq and the compatibility condition
for $j=3$ is valid since \be
(_{III}\mathcal{K}_{1}-3/2\mathcal{I})\mathbf{c}_{3} =\left(
  \begin{array}{cccc}
    -\dfrac{7}{4} & 1 & 0 & 0\\ \\
    -\dfrac{3}{16} & -\dfrac{3}{4} & -2A\lambda c \a & -2A\a\\ \\
    \dfrac{c}{\a} & -4\dfrac{c}{\a} & -\dfrac{3}{2} & 0\\ \\
    \dfrac{3}{16\a A}-\dfrac{\lambda c^{2}}{\a} & -\dfrac{3}{4\a A}+\dfrac{4\lambda c^{2}}{\a} & 0 & -\dfrac{3}{2}\\
  \end{array}
\right) \left(
  \begin{array}{c}
    -\dfrac{\a}{4c}c_{33} \\ \\
    -\dfrac{7\a}{16c}c_{33} \\ \\
     c_{33}\\ \\
     \dfrac{2}{c}\zeta c_{33}\\
  \end{array}
\right)\\
=\left(
          \begin{array}{c}
            0 \\ \\
            0 \\ \\
            0\\ \\
            0 \\
          \end{array}
        \right),
\ee where $\zeta=3/(32A)-\lambda c^{2}/2$. We thus conclude that as
$\Upsilon\rightarrow 0$, Eqs.
(\ref{puiseux_III_1_1_i})-(\ref{puiseux_III_1_1_iv}) describe the
asymptotics around a collapse I singularity with the density of the
fluid diverging there, i.e., \be a\rightarrow 0, \quad a'\rightarrow
\infty, \quad \phi'\rightarrow \infty, \quad \rho_{2}\rightarrow \infty.
\ee
\subsubsection{The balance $_{III}\mathcal{B}_{2}$}
Finally, the balance $_{III}\mathcal{B}_{2}$ corresponds to a
potentially general solution of a flat brane and has the following
$\mathcal{K}$-exponents
\be
\label{speck_III_2}
\textrm{spec}(_{III}\mathcal{K}_{2})=\{-1,0,-2(p-1),1-4p\}.
\ee
The last two $\mathcal{K}$-exponents are positive when $\gamma<-1$ or
$\gamma >1$. We consider here the case $\gamma >1$ because as we
show below it leads to the emergence of a collapse I singularity
while the case $\gamma <-1$ implies the existence of a big rip
singularity and it is considered later in Subsection 3.6. We set $\gamma
=2$. Then $p=1/6$ for which the balance and the
$\mathcal{K}$-exponents read \be
_{III}\mathcal{B}_{2}=\{(\a,\a/6,0,1/(24 A)),(1/6,-5/6,-1,-2)\} \ee
and \be \textrm{spec}(_{III}\mathcal{K}_{2})=\{-1,0,5/3,1/3\}. \ee
Since we have three non-negative $\mathcal{K}$-exponents we see that
this balance indeed corresponds to a general solution. The variables
in this case expand as follows, \bq \label{puiseux_III_2_2_i}
x&=&\a\Upsilon^{1/6}+3/5\a A\lambda c_{13}^{2}\Upsilon ^{5/6}- 27/14
\a A^{2}\lambda^{2}c_{13}^{4}\Upsilon^{3/2}-2\a A c_{54}\Upsilon
^{11/6}
+\ldots\\
y&=&\a/6\Upsilon^{-5/6}+1/2\a A \lambda c_{13}^{2}\Upsilon^{-1/6}-
81/28\a A^{2}\lambda ^{2}c_{13}^{4}\Upsilon^{1/2}-11/3\a A
c_{54}\Upsilon^{5/6}
+\ldots \quad\quad\quad\\
z&=&c_{13}\Upsilon^{-2/3}-12/5 A \lambda c_{13}^{3}+
396/35 A^{2}\lambda^{2}c_{13}^{5}\Upsilon^{2/3}+\ldots\\
\label{puiseux_III_2_2_iv} w&=&1/(24A)\Upsilon^{-2}-3/10\lambda
c_{13}^{2}\Upsilon^{-4/3}+ 747/350 A \lambda^{2}c_{13}^{4}\Upsilon
^{-2/3}+c_{54}\Upsilon^{-1/3}+\ldots \eq We ought to check the
compatibility conditions for $j=1$ and $j=5$. We find \be
(_{III}\mathcal{K}_{2}-1/3\mathcal{I})\mathbf{c}_{1}= \left(
  \begin{array}{cccc}
    -1/2       & 1   & 0 & 0\\
    -5/36      & 1/2 & 0 & -10/3\a A\\
     0 & 0     & 0 & 0\\
    1/(12\a A) & -1/(2\a A) & 0 & -1/3\\
  \end{array}
\right) \left(
  \begin{array}{c}
     0 \\
     0 \\
     c_{13}\\
     0\\
  \end{array}
\right)=\left(
          \begin{array}{c}
            0 \\
            0 \\
            0 \\
            0 \\
          \end{array}
        \right)
\ee so that the compatibility condition for $j=1$ is satisfied. Also
for $j=5$ we have \bq \nonumber
(_{III}\mathcal{K}_{2}-5/3\mathcal{I})\mathbf{c}_{5}&=& \left(
  \begin{array}{cccc}
    -11/6      &  1         & 0    & 0\\
     -5/36     & -5/6       & 0    & -10/3\a A\\
     0         &  0         & -4/3 & 0\\
    1/(12\a A) & -1/(2\a A) & 0    & -5/3\\
  \end{array}
\right) \left(
  \begin{array}{c}
    -2\a A c_{54} \\
    -11/3 \a A c_{54} \\
    c_{53}\\
    c_{54}\\
  \end{array}
\right)=\\
&=&\left(
          \begin{array}{c}
            0 \\
            0 \\
            -4/3 c_{53}\\
            0 \\
          \end{array}
        \right)=P_{5},
\eq where $c_{53}=396/35 A^{2}\lambda^{2}c_{13}^{5}$. The
corresponding eigenvector $v$ is such that
$$v^{\top}=(1/(12 A \a),-1/(2 A \a), 0,1).$$
The compatibility condition, \be v^{\top}\cdot P_{j}=0, \ee for $j=5$
therefore holds true and Eqs.
(\ref{puiseux_III_2_2_i})-(\ref{puiseux_III_2_2_iv}) represent the
asymptotics around a collapse I singularity with the density of the
fluid being divergent, i.e. as $\Upsilon\rightarrow 0$
\be
a\rightarrow 0, \quad a'\rightarrow \infty, \quad \phi'\rightarrow
\infty, \quad \rho_{2}\rightarrow \infty. \ee
\subsection{Collapse type II singularity}
We shall analyze in this Section the asymptotics represented by the
balances $_{II}\mathcal{B}_{3}$ and $_{III}\mathcal{B}_{4}$ which suggest the
emergence of a collapse type II singularity.

We start with the balance $_{II}\mathcal{B}_{3}$ that corresponds to a curved brane.
The $\mathcal{K}$-matrix for this balance is given by
\be
_{II}\mathcal{K}_{3}=D\mathbf{f}^{(0)}(\alpha,\alpha,0,0)
-\textrm{diag}(1,0,r,-2),
\ee
where $D\mathbf{f}^{(0)}$ is the Jacobian matrix of the dominant part
$\mathbf{f}^{(0)}$ in Eq. (\ref{dominantpart2}). The eigenvalues of $_{II}\mathcal{K}_{3}$ are
\be
\textrm{spec}(_{II}\mathcal{K}_{3})=\{-1,0,-2-4\gamma,-4-r\}.
\ee
For $\gamma=-1$ and $r=1$, this balance becomes
\be
_{II}\mathcal{B}_{3}=\{(\a,\a,0,0),(1,0,1,-2)\},
\ee
and the eigenvalues of the $\mathcal{K}$-matrix are
\be
\textrm{spec}(_{II}\mathcal{K}_{3})=\{-1,0,2,-5\}.
\ee
We may set the arbitrary constant appearing for $j=-5$ equal to zero
and find a particular solution 
around a collapse II singularity.
The asymptotic expansions of the variables now are:
\bq
\label{puiseux_II_3_-1_i}
x&=&\a\Upsilon+A\a/9c_{24}\Upsilon^{3}+\ldots\\
y&=&\a+A\a/3c_{24}\Upsilon^{2}+\ldots\\
z&=&0+\ldots\\
\label{puiseux_II_3_-1_iv}
w&=&c_{24}+\ldots \, .
\eq

The compatibility condition for $j=2$ is satisfied since
 \be
 (_{II}\mathcal{K}_{3}-2\mathcal{I})\mathbf{c}_{2}=
\left(
  \begin{array}{cccc}
   -3 & 1 & 0 & 0\\
    0 & -2 & 0 & (2/3)\a A\\
    0 &  0 & -3 & 0\\
    0 &  0 & 0 & 0\\
  \end{array}
\right) \left(
  \begin{array}{c}
    (\a A/9)c_{24} \\
    (\a A/3)c_{24} \\
    0\\
    c_{24}\\
  \end{array}
\right)=\left(
          \begin{array}{c}
            0 \\
            0 \\
            0 \\
            0 \\
          \end{array}
        \right).
\ee
It follows then from Eqs.
(\ref{puiseux_II_3_-1_i})-(\ref{puiseux_II_3_-1_iv}) that as
$\Upsilon\rightarrow 0$,
\be
a\rightarrow 0, \quad a'\rightarrow
\a,\quad \phi'\rightarrow 0, \quad \rho_{2}\rightarrow c_{24}.
\ee
%
%
Next we turn to the balance $_{III}\mathcal{B}_{4}$. This balance
corresponds to a potentially general solution of a curved brane.
The eigenvalues of its $\mathcal{K}$-matrix are
\be \label{spec_K_III_4}
\textrm{spec}(_{III}\mathcal{K}_{4})=\{-3,-2-4\gamma,-1,0\}.
\ee
For $\gamma <-1/2$ we have one positive eigenvalue (the case $\gamma
>-1/2$ for which we have three negative eigenvalues is considered
later in Subsection 3.7). We may set $\mathbf{c}_{-3}=0$ from the beginning and
find an asymptotic expansion of a particular solution
around a collapse II singularity. We choose $\gamma=-3/4$. Then
\be
\textrm{spec}(_{III}\mathcal{K}_{4})=\{-3,1,-1,0\},
\ee
and we find the following asymptotic forms
\bq
\label{puiseux_3_4_-3/4_i}
x&=&\a\Upsilon+A\a/6 c_{14}\Upsilon ^{2}+\ldots\\
y&=&\a+A\a/3c_{14}\Upsilon+\ldots\\
z&=&0+\ldots\\
\label{puiseux_3_4_-3/4_iv} w&=&c_{14}\Upsilon^{-1}+\ldots.
\eq
The compatibility condition for $j=1$ is trivially satisfied since
\be
(_{III}\mathcal{K}_{4}-\mathcal{I})\mathbf{c}_{1}= \left(
  \begin{array}{cccc}
    -2 & 1    & 0  & 0\\
     0 & -1   & 0  & \a A/3\\
     0 & 0    & -4 & 0\\
     0 & 0    & 0  & 0\\
  \end{array}
\right) \left(
  \begin{array}{c}
    (\a A/6) c_{14} \\
    (\a A/3)c_{14} \\
    0\\
    c_{14}\\
  \end{array}
\right)=\left(
          \begin{array}{c}
            0 \\
            0 \\
            0\\
            0 \\
          \end{array}
        \right).
\ee
The forms in Eqs. (\ref{puiseux_3_4_-3/4_i})-(\ref{puiseux_3_4_-3/4_iv}) then imply that
as $\Upsilon\rightarrow 0$,
\be
a\rightarrow 0, \quad a'\rightarrow
\a, \quad \phi'\rightarrow 0, \quad \rho_{2}\rightarrow \infty.
\ee

Thus for a curved brane we found collapse II singularities with
$\phi'\rightarrow 0$ and $\rho_{2}\rightarrow \infty, \rho_{s}$, $\rho_{s}\neq 0$.
This means that asymptotically the leak of energy
from the brane and into the bulk is controlled solely by the fluid.
\subsection{Big rip singularity}
In this Section we examine the balances $_{II}\mathcal{B}_{1}$ and
$_{III}\mathcal{B}_{2}$ which correspond to a flat brane and describe the asymptotics around a big
rip singularity when we choose a value of $\gamma$ such that
$\gamma<-1$.

The first of these balances has the following $\mathcal{K}$-exponents
\be
\textrm{spec}(_{II}\mathcal{K}_{1})=\{-1,0,0,-2(p-1)\}.
\ee
Choosing $\gamma=-2$, we have that
\be
_{II}\mathcal{B}_{1}=\{(\a,-\a/2,c,3/(8A)),(-1/2,-3/2,2,-2)\}
\ee
and 
\be _{II}\mathcal{K}_{1}=\{-1,0,0,3\}. \ee This leads to the
following solution, \bq
x&=&\a\Upsilon^{-1/2}-\a/(4c) c_{33}\Upsilon ^{5/2}+\ldots\\
y&=&-\a/2\Upsilon^{-3/2}-5\a/(8c)c_{33}\Upsilon^{3/2}+\ldots\\
z&=&c\Upsilon^{2}+c_{33}\Upsilon^{5}+\ldots\\
w&=&3/(8A)\Upsilon^{-2}-3/(8Ac)c_{33}\Upsilon+\ldots \eq We validate
the compatibility condition for $j=3$, since \be
(_{II}\mathcal{K}_{1}-3\mathcal{I})\mathbf{c}_{3}= \left(
  \begin{array}{cccc}
    -5/2 & 1 & 0 & 0\\
    3/4 & -3/2 & 0 & 2\a A\\
    -2c/\a & -4c/\a & -3 & 0\\
    3/(4\a A) &  3/(2\a A) & 0 & -3\\
  \end{array}
\right) \left(
  \begin{array}{c}
    -\a/(4c) c_{33} \\
    -5\a/(8c)c_{33} \\
    c_{33}\\
    -3/(8A c)c_{33}\\
  \end{array}
\right)=\left(
          \begin{array}{c}
            0 \\
            0 \\
            0\\
            0 \\
          \end{array}
        \right).
\ee As $\Upsilon\rightarrow 0$ we have a big rip singularity: \be
a\rightarrow \infty, \quad a'\rightarrow -\infty, \quad
\phi'\rightarrow 0, \quad \rho_{2}\rightarrow \infty. \ee

We now examine the second balance $_{III}\mathcal{B}_{2}$. The
$\mathcal{K}$-exponents for this balance are given by Eq. (\ref{speck_III_2}). We are
interested in the case $\gamma<-1$ which we
have not examined yet and for which we get two positive
$\mathcal{K}$-exponents. As we show below this case leads to the
emergence of a big rip singularity. We set for concreteness
$\gamma=-3/2$. Then \be
_{III}\mathcal{B}_{2}=\{(\a,-\a,0,3/(2A)),(-1,-2,-1,-2)\} \ee and
\be _{III}\mathcal{K}_{2}=\{-1,0,4,5\}. \ee In this case we have
three non-negative $\mathcal{K}$-exponents so that this balance
indeed corresponds to a general solution. The variables in this case
expand as follows, \bq \label{puiseux_III_2_-3/2_i}
x&=&\a\Upsilon^{-1}+A\a/3 c_{44}\Upsilon ^{3}+\ldots\\
y&=&-\a\Upsilon^{-2}+A\a c_{44}\Upsilon^{2}+\ldots\\
z&=&c_{53}\Upsilon^{4}+\ldots\\
\label{puiseux_III_2_-3/2_iv}
w&=&3/(2A)\Upsilon^{-2}+c_{44}\Upsilon^{2}+\ldots \eq We ought to
check the compatibility conditions for $j=4$ and $j=5$. We find \be
(_{III}\mathcal{K}_{2}-4\mathcal{I})\mathbf{c}_{4}= \left(
  \begin{array}{cccc}
    -3 & 1  & 0 & 0\\
     2 & -2 & 0 & (4/3)\a A\\
     0 & 0  & -1 & 0\\
    3/(\a A) &  3/(\a A) & 0 & -4\\
  \end{array}
\right) \left(
  \begin{array}{c}
     A\a/3 c_{44} \\
     A\a c_{44} \\
     0\\
     c_{44}\\
  \end{array}
\right)=\left(
          \begin{array}{c}
            0 \\
            0 \\
            0\\
            0 \\
          \end{array}
        \right)
\ee and \be (_{III}\mathcal{K}_{2}-5\mathcal{I})\mathbf{c}_{5}=
\left(
  \begin{array}{cccc}
    -4 &  1 & 0 & 0\\
     2 & -3 & 0 & (4/3)\a A\\
     0 &  0 & 0 & 0\\
    3/(\a A) & 3/(\a A) & 0 & -5\\
  \end{array}
\right) \left(
  \begin{array}{c}
    0 \\
    0 \\
    c_{53}\\
    0\\
  \end{array}
\right)=\left(
          \begin{array}{c}
            0 \\
            0 \\
            0\\
            0 \\
          \end{array}
        \right),
\ee so that Eqs.
(\ref{puiseux_III_2_-3/2_i})-(\ref{puiseux_III_2_-3/2_iv}) represent
the asymptotics around a big rip singularity i.e. as
$\Upsilon\rightarrow 0$ \be a\rightarrow \infty, \quad a'\rightarrow
-\infty, \quad \phi'\rightarrow 0, \quad \rho_{2}\rightarrow \infty. \ee

Hence in both of the cases studied in this Section we found that for a flat brane, a big rip
singularity develops at a finite distance. We note here that the exchange
of energy from the brane into the bulk is totally monitored by the fluid.
\subsection{Behavior at infinity}
In this Section we consider the balances $_{I}\mathcal{B}_{3}$,
$_{III}\mathcal{B}_{4}$, $_{I}\mathcal{B}_{4}$,
$_{II}\mathcal{B}_{2}$ and $_{III}\mathcal{B}_{2}$ that offer the
possibility of escaping the finite-distance singularities met before
and describe the behavior of our model at infinity.

We begin with the analysis of the balance $_{I}\mathcal{B}_{3}$.
The eigenvalues of its $\mathcal{K}$-matrix read
\be
\textrm{spec}(_{I}\mathcal{K}_{3})=\{-1,-3,0,0\},
\ee
hence we may expand $(x,y,z,w)$ in descending powers in order to meet the
arbitrary constants appearing at $j=-1$ and $j=-3$. We choose
$\gamma=-1/2$. The balance $_{I}\mathcal{B}_{3}$ then
corresponds to a 
general solution of a flat or curved brane. In particular we find
\bq
\label{puiseux_1_3_-1/2_i}
x&=&\a\Upsilon+c_{-1 1}+\ldots\\
y&=&\a+\ldots\\
z&=&c_{-3 3}\Upsilon^{-4}+\ldots\\
\label{puiseux_1_3_-1/2_iv} w&=&\zeta\Upsilon^{-2}-2\zeta/\a c_{-1
1}\Upsilon^{-3}+ 3\zeta/\a^{2} c_{-1
1}^{2}\Upsilon^{-4}-4\zeta/\a^{3}c_{-1 1}^{3}\Upsilon^{-5}+\ldots .
\eq
The compatibility conditions for $j=-1$ are satisfied since \be
(_{I}\mathcal{K}_{3}+\mathcal{I})\mathbf{c}_{-1}= \left(
  \begin{array}{cccc}
     0         & 1          & 0    & 0\\
     0         & 1          & 0    & 0\\
     0         & 0          & -2   & 0\\
     2\zeta/\a & -2\zeta/\a & 0    & 1\\
  \end{array}
\right) \left(
  \begin{array}{c}
     c_{-11} \\
     0 \\
     0\\
     -2\zeta/\a c_{-11}\\
  \end{array}
\right)=\left(
          \begin{array}{c}
            0 \\
            0 \\
            0\\
            0 \\
          \end{array}
        \right).
\ee For $j=-3$ we find \be
(_{I}\mathcal{K}_{3}+3\mathcal{I})\mathbf{c}_{-3}= \left(
  \begin{array}{cccc}
     2         & 1        & 0 & 0\\
     0         & 3        & 0 & 0\\
     0         & 0        & 0 & 0\\
     2\zeta/\a & -2\zeta/\a & 0 & 3\\
  \end{array}
\right) \left(
  \begin{array}{c}
     0 \\
     0 \\
     c_{-33}\\
     -4\zeta/\a^{3} c_{-1 1}^{3}\\
  \end{array}
\right)=\left(
          \begin{array}{c}
            0 \\
            0 \\
            0\\
            -12\zeta/\a^{3} c_{-1 1}^{3}\\
          \end{array}
        \right)=P_{-3}.
\ee Since a corresponding eigenvector here is $v^{\top}=(0,0,1,0)$
and \be v^{\top}\cdot P_{-3}=0 \ee the compatibility condition for
$j=-3$ is also satisfied and Eqs.
(\ref{puiseux_1_3_-1/2_i})-(\ref{puiseux_1_3_-1/2_iv}) then imply
that as $\Upsilon\rightarrow\infty$ \be a\rightarrow \infty, \quad
a'\rightarrow\a, \quad \phi'\rightarrow 0, \quad\rho_{2}\rightarrow 0.
\ee

We now examine the balance 
$ _{III}\mathcal{B}_{4}=\{(\a,\a,0,0),(1,0,-1,-2)\} $ that
corresponds to a general solution of a curved brane. The
$\mathcal{K}$-exponents are given by Eq. (\ref{spec_K_III_4}). For $\gamma
>-1/2$ we have three negative $\mathcal{K}$-exponents. We choose
$\gamma=0$. Then \be
\textrm{spec}(_{III}\mathcal{K}_{4})=\{-3,-2,-1,0\}.
\ee
We find here that
\bq
\label{puiseux_I_4_0_i}
x&=&\a\Upsilon+c_{-1 1}-(1/3)\a A c_{-24}\Upsilon^{-1}-(5/9)Ac_{-1 1}c_{-2 4}\Upsilon^{-2}+\ldots\\
y&=&\a+(1/3)\a A c_{-2 4}\Upsilon^{-2} +(10/9)A c_{-1 1}c_{-2 4}\Upsilon^{-3}+\ldots\\
z&=&c_{-3 3}\Upsilon^{-4}+\ldots\\
\label{puiseux_I_4_0_iv}
w&=&c_{-24}\Upsilon^{-4}+(4/\a)c_{-11}c_{-24}\Upsilon^{-5}+\ldots\, .
\eq
We check the compatibility conditions for $j=-1$, $j=-2$ and
$j=-3$. For $j=-1$ we find \be
(_{III}\mathcal{K}_{4}+\mathcal{I})\mathbf{c}_{-1}= \left(
  \begin{array}{cccc}
     0 & 1 & 0 & 0\\
     0 & 1 & 0 & -(2/3)\a A\\
     0 & 0 & -2 & 0\\
     0 & 0 & 0 & -1\\
  \end{array}
\right) \left(
  \begin{array}{c}
     c_{-11} \\
     0 \\
     0\\
     0\\
  \end{array}
\right)=\left(
          \begin{array}{c}
            0 \\
            0 \\
            0\\
            0 \\
          \end{array}
        \right),
\ee while for $j=-2$ \be
(_{III}\mathcal{K}_{4}+2\mathcal{I})\mathbf{c}_{-2}= \left(
  \begin{array}{cccc}
     1 & 1 & 0 & 0\\
     0 & 2 & 0 & -(2/3)\a A\\
     0 & 0 & -1 & 0\\
     0 & 0 & 0 & 0\\
  \end{array}
\right) \left(
  \begin{array}{c}
    -(1/3)\a A c _{-24} \\
    (1/3)\a A c_{-24} \\
     0\\
     c_{-24}\\
  \end{array}
\right)=\left(
          \begin{array}{c}
            0 \\
            0 \\
            0\\
            0 \\
          \end{array}
        \right).
\ee For $j=-3$ we have that \be \nonumber
(_{III}\mathcal{K}_{4}+3\mathcal{I})\mathbf{c}_{-3}= \left(
  \begin{array}{cccc}
     2 &  1  & 0 & 0\\
     0 &  3  & 0 & -(2/3)\a A\\
     0 &  0  & 0 & 0\\
     0 &  0  & 0 & 1\\
  \end{array}
\right) \left(
  \begin{array}{c}
  -(5/9) A c_{-11}c_{-24} \\
  (10/9) A c_{-11}c_{-24} \\
  c_{-33}\\
  (4/\a) c_{-11}c_{-24}\\
  \end{array}
\right)= \ee \be =\left(
          \begin{array}{c}
            0 \\
            2/3 Ac_{-11}c_{-24} \\
            0\\
            4/\a c_{-11}c_{-24} \\
          \end{array}
        \right)=P_{-3},
\ee and an eigenvector $v$ is such that $v^{\top}=(0,0,1,0)$. The
compatibility condition, \be v^{\top}\cdot P_{-3}=0, \ee therefore
holds true. Eqs. (\ref{puiseux_I_4_0_i})-(\ref{puiseux_I_4_0_iv})
then imply that as $\Upsilon\rightarrow\infty$
\be
a\rightarrow \infty, \quad a'\rightarrow\a, \quad \phi'\rightarrow
0, \quad\rho_{2}\rightarrow 0. \ee

On the other hand, the $\mathcal{K}$-exponents for the balance $_{I}\mathcal{B}_{4}$,
that also corresponds to a curved brane, are
\be
\textrm{spec}(_{I}\mathcal{K}_{4})=\{-3,-s-4(\gamma+1),-1,0\}.
\ee
For $s=-3$ and $\gamma=1/4$ we find the following behavior
\bq
\label{puiseux_I_4_1/4_i}
x&=&\a\Upsilon+c_{-1 1}-(1/6)\a A c_{-24}\Upsilon^{-2}+\ldots\\
y&=&\a+(1/3)\a A c_{-2 4}\Upsilon^{-3}+\ldots\\
z&=&c_{-3 3}\Upsilon^{-4}+\ldots\\
\label{puiseux_I_4_1/4_iv}
w&=&c_{-24}\Upsilon^{-5}-(5/\a)c_{-11}c_{-24}\Upsilon^{-6}+\ldots\, .
\eq
The compatibility conditions for $j=-1$ and $j=-2$ are trivially
satisfied since
\be
(_{I}\mathcal{K}_{4}+\mathcal{I})\mathbf{c}_{-1}= \left(
  \begin{array}{cccc}
     0 & 1 & 0 & 0\\
     0 & 1 & 0 & 0\\
     0 & 0 & -2 & 0\\
     0 & 0 & 0 & -1\\
  \end{array}
\right) \left(
  \begin{array}{c}
     c_{-11} \\
     0 \\
     0\\
     0\\
  \end{array}
\right)=\left(
          \begin{array}{c}
            0 \\
            0 \\
            0\\
            0 \\
          \end{array}
        \right)
\ee
and
\be
(_{I}\mathcal{K}_{4}+2\mathcal{I})\mathbf{c}_{-2}= \left(
  \begin{array}{cccc}
     1 & 1 & 0 & 0\\
     0 & 2 & 0 & 0\\
     0 & 0 & -1 & 0\\
     0 & 0 & 0 & 0\\
  \end{array}
\right) \left(
  \begin{array}{c}
    0 \\
    0 \\
    0\\
    c_{-24}\\
  \end{array}
\right)=\left(
          \begin{array}{c}
            0 \\
            0 \\
            0\\
            0 \\
          \end{array}
        \right).
\ee
For $j=-3$ we have that
\be \nonumber
(_{I}\mathcal{K}_{4}+3\mathcal{I})\mathbf{c}_{-3}= \left(
  \begin{array}{cccc}
     2 &  1  & 0 & 0\\
     0 &  3  & 0 & 0\\
     0 &  0  & 0 & 0\\
     0 &  0  & 0 & 1\\
  \end{array}
\right) \left(
  \begin{array}{c}
  -(1/6) A \a c_{-24} \\
  (1/3) A \a c_{-24} \\
  c_{-33}\\
  -(5/\a) c_{-11}c_{-24}\\
  \end{array}
\right)= \ee \be =\left(
          \begin{array}{c}
            0 \\
            A \a c_{-24} \\
            0\\
            -(5/\a) c_{-11}c_{-24} \\
          \end{array}
        \right)=P_{-3},
\ee
and an eigenvector $v$ is such that $v^{\top}=(0,0,1,0)$ so that
the compatibility condition for $j=-3$,
\be
v^{\top}\cdot P_{-3}=0,
\ee
is also satisfied.
Therefore Eqs. (\ref{puiseux_I_4_1/4_i})-(\ref{puiseux_I_4_1/4_iv})
then imply that as $\Upsilon\rightarrow\infty$
\be
a\rightarrow \infty, \quad a'\rightarrow\a, \quad \phi'\rightarrow
0, \quad\rho_{2}\rightarrow 0.
\ee

Let us now examine the balance $_{II}\mathcal{B}_{2}$ that corresponds to a flat brane. For
$\gamma=-3/4$ (hence $p=2$) and $r=-5$ this balance reads
$_{II}\mathcal{B}_{2}=\{(\a,2\a,0,6/A),(2,1,-5,-2)\}$. and the
eigenvalues of the $_{II}\mathcal{K}_{2}$ matrix are
\be
\textrm{spec}(_{II}\mathcal{K}_{2})=\{-3,-2,-1,0\}.
\ee
We find here the following asymptotic behavior
\bq
\label{puiseux_2_2_-3/4_i}
x&=&\a\Upsilon^{2}-A\a/6c_{-1\,4}\Upsilon-A\a/6c_{-2\,4}+
A^{2}\a/36c_{-1\,4}^{2}+\ldots \\
y&=&2\a\Upsilon-A\a/6c_{-1\,4}+\ldots\\
z&=&c_{-3\,3}\Upsilon^{-8}+\ldots\\
\label{puiseux_2_2_-3/4_iv}
w&=&6/A\Upsilon^{-2}+c_{-1\,4}\Upsilon^{-3}+c_{-2\,4}\Upsilon^{-4}+
(-A^{2}/36c_{-1\,4}^{3}+A/3c_{-1\,4}c_{-2\,4})\Upsilon^{-5}+\ldots .
\eq
The compatibility conditions for $j=-1$, $j=-2$ and $j=-3$ are satisfied. Particularly for
$j=-1$ we find \be
(_{II}\mathcal{K}_{2}+\mathcal{I})\mathbf{c}_{-1}= \left(
  \begin{array}{cccc}
     -1        & 1        & 0  & 0\\
     2         & 0        & 0  & A\a/3\\
     0         & 0        & -2 & 0\\
     12/(A\a)  & -6/(A\a) & 0  & 1\\
  \end{array}
\right) \left(
  \begin{array}{c}
     -A\a/6c_{-1\,4} \\
     -A\a/6c_{-1\,4} \\
     0\\
     c_{-1\,4}\\
  \end{array}
\right)=\left(
          \begin{array}{c}
            0 \\
            0 \\
            0\\
            0 \\
          \end{array}
        \right).
\ee
For $j=-2$ we find that
\be
\nonumber
(_{II}\mathcal{K}_{2}+2\mathcal{I})\mathbf{c}_{-2}= \left(
  \begin{array}{cccc}
     0         & 1        & 0  & 0\\
     2         & 1        & 0  & A\a/3\\
     0         & 0        & -1 & 0\\
     12/(A\a)  & -6/(A\a) & 0  & 2\\
  \end{array}
\right) \left(
  \begin{array}{c}
     -A\a/6c_{-2\,4}+A^{2}\a/36c_{-1\,4}^{2} \\
     0 \\
     0\\
     c_{-2\,4}\\
  \end{array}
\right)= \ee \be =\left(
          \begin{array}{c}
            0 \\
            A^{2}\a/18c_{-1\,4}^{2} \\
            0\\
            A/3c_{-1\,4}^{2} \\
          \end{array}
        \right)=P_{-2},
\ee
and an eigenvector $v$ is such that
\be
v^{\top}=(12/(A \a),-6/(A \a),0,1)
\ee
and hence we find
\be
v^{\top}\cdot P_{-2}=0,
\ee
which means that the compatibility condition is satisfied also for $j=-2$.
Finally, for $j=-3$ we see that \be \nonumber
(_{II}\mathcal{K}_{2}+3\mathcal{I})\mathbf{c}_{-3}= \left(
  \begin{array}{cccc}
     1         & 1        & 0 & 0\\
     2         & 2        & 0 & A\a/3\\
     0         & 0        & 0 & 0\\
     12/(A\a)  & -6/(A\a) & 0 & 3\\
  \end{array}
\right) \left(
  \begin{array}{c}
     0 \\
     0 \\
     c_{-3\,3}\\
     -A^{2}/36c_{-1\,4}^{3}+A/3c_{-1\,4}c_{-2\,4}\\
  \end{array}
\right)= \ee \be =\left(
          \begin{array}{c}
            0 \\
            A\a/3(-A^{2}/36c_{-1\,4}^{3}+A/3c_{-1\,4}c_{-2\,4}) \\
            0\\
            3 (-A^{2}/36c_{-1\,4}^{3}+A/3c_{-1\,4}c_{-2\,4}) \\
          \end{array}
        \right)=P_{-3}
\ee while an eigenvector $v$ is such that
\be
v^{\top}=(0,0,1,0),
\ee
and hence
\be
v^{\top}\cdot P_{-3}=0, \ee so that the compatibility condition
for $j=-3$ is satisfied. We thus see from Eqs.
(\ref{puiseux_2_2_-3/4_i})-(\ref{puiseux_2_2_-3/4_iv}) that as
$\Upsilon\rightarrow\infty$ \be a\rightarrow \infty, \quad
a'\rightarrow\infty, \quad \phi'\rightarrow 0,
\quad\rho_{2}\rightarrow 0. \ee

We now move on to the balance $_{III}\mathcal{B}_{2}$ which also
corresponds to a flat brane. It follows from (\ref{speck_III_2}) that for $-1<\gamma<-1/2$
this balance leads to three negative $\mathcal{K}$-exponents. We take
$\gamma=-3/4$ and then this balance becomes
\be _{III}\mathcal{B}_{2}=\{(\a,2\a,0,6/A),(2,1,-1,-2)\}, \ee and
the eigenvalues of the $_{III}\mathcal{K}_{2}$ matrix are \be
\textrm{spec}(_{III}\mathcal{K}_{2})=\{-7,-2,-1,0\}. \ee
In this case we find that
\bq
\label{puiseux_3_2_-3/4_i}
x&=&\a\Upsilon^{2}+c_{-1\,1}\Upsilon+c_{-2\,1}+\ldots\\
y&=&2\a\Upsilon+c_{-1\,1}+\ldots\\
z&=&c_{-7\,3}\Upsilon^{-8}+\ldots\\
\label{puiseux_3_2_-3/4_iv}
\nonumber
w&=&6/A\Upsilon^{-2}-6/(A \a)c_{-1\,1}\Upsilon^{-3}+
6/(A\a^{2})(c_{-1\,1}^{2}-\a c_{-2\,1})\Upsilon^{-4}+\\
&+&c_{-3\,4}\Upsilon^{-5}+c_{-4\,4}\Upsilon^{-6}+c_{-5\,4}\Upsilon^{-7}+
c_{-6\,4}\Upsilon^{-8}+c_{-7\,4}\Upsilon^{-9}+\ldots,
\eq
where $c_{j\,4}$ with $j=-3,\ldots,-7$, are polynomials in $\a$, $c_{-1\,1}$ and $c_{-2\, 1}$.
The compatibility condition at $j=-1$ is trivially satisfied since
we have that
\be
(_{III}\mathcal{K}_{2}+\mathcal{I})\mathbf{c}_{-1}= \left(
  \begin{array}{cccc}
    -1        & 1        & 0  & 0\\
     2         & 0        & 0  & A\a/3\\
     0         & 0        & -6 & 0\\
     12/(A\a)  & -6/(A\a) & 0  & 1\\
  \end{array}
\right) \left(
  \begin{array}{c}
     c_{-1\,1} \\
     c_{-1\,1} \\
     0\\
     -6/(A\a)c_{-1\,1}\\
  \end{array}
\right)=\left(
          \begin{array}{c}
            0 \\
            0 \\
            0\\
            0 \\
          \end{array}
        \right).
\ee
For $j=-2$ we find that
\be
(_{III}\mathcal{K}_{2}+2\mathcal{I})\mathbf{c}_{-2}= \left(
  \begin{array}{cccc}
     0         & 1        & 0  & 0\\
     2         & 1        & 0  & A\a/3\\
     0         & 0        & -5 & 0\\
     12/(A\a)  & -6/(A\a) & 0  & 2\\
  \end{array}
\right) \left(
          \begin{array}{c}
            c_{-2\,1} \\
             0 \\
             0\\
             6/(A\a^{2})(c_{-1\,1}^{2}-\a c_{-2\,1}) \\
          \end{array}\right)=
\ee
\be
    =\left(
          \begin{array}{c}
            0 \\
            2/\a c_{-1\,1}^{2} \\
            0\\
            12/(A \a^{2})c_{-1\,1}^{2} \\
          \end{array}
        \right)=P_{-2},
\ee
and $v^{\top}=(12/(A \a),-6/(A \a),0,1)$
so that the compatibility condition $$v^{\top}\cdot P_{-2}=0$$
is satisfied. Finally, for $j=-7$ we have that
\be (_{III}\mathcal{K}_{2}+7\mathcal{I})\mathbf{c}_{-7}=
\left(
  \begin{array}{cccc}
     5         & 1        & 0  & 0\\
     2         & 6        & 0  & A\a/3\\
     0         & 0        & 0  & 0\\
     12/(A\a)  & -6/(A\a) & 0  & 7\\
  \end{array}
\right) \left(
  \begin{array}{c}
     0 \\
     0 \\
     c_{-7\,3}\\
     c_{-7\,4}\\
  \end{array}
\right)=\left(
          \begin{array}{c}
            0 \\
            A\a/3c_{-7\,4} \\
            0\\
            7c_{-7\,4} \\
          \end{array}
        \right)=P_{-7},
\ee and the corresponding eigenvector here is such that
$v^{\top}=(0,0,1,0)$ which implies that the compatibility
condition $$v^{\top}\cdot P_{-7}=0$$ holds true. From Eqs.
(\ref{puiseux_3_2_-3/4_i})-(\ref{puiseux_3_2_-3/4_iv}) it follows
that as $\Upsilon\rightarrow\infty$
\be
a\rightarrow \infty, \quad a'\rightarrow\infty, \quad
\phi'\rightarrow 0, \quad\rho_{2}\rightarrow 0. \ee

In our previous work in \cite{ack}, the bulk was filled entirely by
the fluid and we found that for $-1<\gamma<-1/2$ and for a flat brane the avoidance
of finite-distance singularities was the \emph{only} possible asymptotic behavior. This fact
suggested that a self-tuning mechanism could be build within the framework of such model,
a property that we would anticipate to hold also in the more complicated case studied
in this paper. However, by the analysis we did so far, we see instead that although the
balance $_{III}\mathcal{B}_{2}$ implies a behavior that is singular only at infinite
distance, the balance $_{III}\mathcal{B}_{3}$ that is also valid for this range of $\gamma$
implies a singular (collapse type I) behavior at finite distance (as this follows from
Eqs. (\ref{puisuex_3_3_-3/4_i})-(\ref{puiseux_3_3_-3/4_iv}) in Subsection 3.4.2.).
Both of these balances represent behaviors of the general solution. In particular,
the balance $_{III}\mathcal{B}_{2}$ describes the asymptotic behavior of the general
solution in a neighborhood of infinity, while, the balance $_{III}\mathcal{B}_{3}$ describes
the asymptotic behavior of the general solution around a finite-distance singularity.
We thus conclude that in this case the avoidance of finite-distance singularities for flat brane
becomes impossible. Our results suggest that the singular behavior encountered here is driven by the
presence of the scalar field which is left to act independently from the fluid. What would
happen, instead, if it interacted with the fluid? In the next Section, we show that by choosing
the interaction parameters in an adequate way, we may resolve this unwanted situation and
recover the possibility of avoiding finite-distance singularities.
\section{Interacting mixture in the bulk}
In this Section,  we study the possible behaviors that arise when the two bulk components
interact with each other. We begin by searching to find what are the forms of the balances
that are possible in this more intricate case. In order to simplify our calculations, that are
much more complicated than in the case studied previously, we may set to zero either one
of the two parameters $\sigma$ and $\nu$ that define the interaction and let the remaining
one vary arbitrarily. If we choose both parameters nonzero we are led to balances
$\{\mathbf{a},\mathbf{p}\}$ with the exponents in the vector $\mathbf{p}$ being irrational
which leads to the existence of logarithms in the series expansions of the variables. In the
next paragraphs we show that the choice that gives the desired result, meaning the
avoidance of singularities, is $\sigma=0$.
\footnote{The case $\nu=0$ and $\sigma$ arbitrary does not lead to avoidance of
singularities but instead it brings back the same problem we faced in Subsection 3.7.}

We start the analysis by putting $\sigma=0$ in the system (\ref{syst1_1_n_s})-(\ref{syst1_4_n_s})
and letting $\nu$ be arbitrary but nonzero. We consider all terms dominant and by substituting
Eq. (\ref{variables}) there, we find the following four balances
\bq
_{\nu}\mathcal{B}_{1}&=&\left\{
\left(\a,\dfrac{2\a}{8+\nu},\sqrt{\frac{3(4\gamma-4-\nu)}{A\lambda (\gamma-1)(8+\nu)^{2}}},
\frac{3\nu}{2A(\gamma-1)(8+\nu)^{2}}\right),
\left(\frac{2}{8+\nu},-\frac{6+\nu}{8+\nu},-1,-2\right)\right\}\quad\quad\,
\\
_{\nu}\mathcal{B}_{2}&=&\left\{\left(\a,\a p,0,
\frac{3p^{2}}{2A}\right), (p,p-1,-1,-2)\right\},\,\, p=\frac{1}{2(\gamma+1)},\,\,\gamma\neq-1,-1/2,
\\
_{\nu}\mathcal{B}_{3}&=&\{(\a,\a,0,0), (1,0,-1,-2)\},
\\
_{\nu}\mathcal{B}_{4}&=&\{(\a,0,0,0),(0,-1,-1,-2)\}.
\eq
The balance $_{\nu}\mathcal{B}_{1}$ is valid for $\nu\neq-6,-8,4\gamma-4$ and
$\gamma\neq1$ and because of the square root we may have either $\nu >4\gamma-4$
and $\gamma <1$, or, $\nu <4\gamma-4$ and $\gamma >1$.
Substitution of these balances in the constraint equation (\ref{constraint}) shows that
$_{\nu}\mathcal{B}_{1}$, $_{\nu}\mathcal{B}_{2}$ and $_{\nu}\mathcal{B}_{4}$ correspond
to a flat brane whereas $_{\nu}\mathcal{B}_{3}$ corresponds to a curved brane with $\a$
satisfying $\a^{2}=kH^{2}$.

We now calculate the Jacobian of the vector field (\ref{vectorfield_full_n_s}) (with $\sigma=0$)
and find
\bq
\nonumber
& & D\mathbf{f}(x,y,z,w)=
\\ \nonumber
&=&\left(
                     \begin{array}{cccc}
0                                                                                                                 & 1                                                                                                 & 0                                                            & 0 \\ \\
-\dfrac{2}{3}A(1 + 2\gamma)w -\lambda A z^{2}                                          & 0                                                                                                 & -2\lambda A z x                                      & -\dfrac{2}{3} A(1+2\gamma)x \\ \\
\left(4+\dfrac{\nu}{2}\right)\dfrac{y z}{x^{2}}                                              & -\left(4+\dfrac{\nu}{2}\right)\dfrac{z}{x}                                     & -\left(4+\dfrac{\nu}{2}\right)\dfrac{y}{x}& 0 \\ \\
4(\gamma +1)\dfrac{y w}{x^{2}}-\dfrac{\nu\lambda}{2}\dfrac{yz^{2}}{x^{2}}& -4(\gamma+1)\dfrac{w}{x}+\dfrac{\nu\lambda}{2}\dfrac{z^{2}}{x}& \nu\lambda\dfrac{y z}{x}                        &-4(\gamma+1)\dfrac{y}{x}\\
                     \end{array}
                   \right).\\
\eq

The balance $_{\nu}\mathcal{B}_{4}$ is discarded because it does not give the $-1$
$\mathcal{K}$-exponent, but it has instead
\be
\textrm{spec}(_{\nu}\mathcal{K}_{4})=\left\{0,1,1,2\right\}.
\ee

The balance $_{\nu}\mathcal{B}_{3}$, on the other hand, has
\be
\textrm{spec}(_{\nu}\mathcal{K}_{3})=\left\{-1,0,-2(1+2\gamma),-3-\nu/2\right\},
\ee
which implies that for a curved brane we may avoid the finite-distance singularities for
$\gamma >-1/2$ and $\nu >-6$. We may also include in this range of $\gamma$ the
value $-1/2$ since then the following balance arises:
 \be
 _{\nu}\mathcal{B}_{-1/2}=\{(\alpha,\alpha,0,\delta),(1,0,-1-2)\}.
 \ee
The balance $_{\nu}\mathcal{B}_{-1/2}$ corresponds to a 
general solution of a flat or curved brane with $\delta=3/(2A)(1-kH^{2}/\alpha^{2})$,
$\delta\neq 0$ and gives
\be
\textrm{spec}(_{\nu}\mathcal{K}_{-1/2})=\{-1,0,0,-3-\nu/2\},
\ee
so that for $\nu>-6$ finite-distance singularities can be avoided.

The $\mathcal{K}$-exponents for the balance $_{\nu}\mathcal{B}_{1}$ are
\be
\textrm{spec}(_{\nu}\mathcal{K}_{1})=\left\{-1,0,\frac{2(6+\nu)}{8+\nu},
\frac{2(4-4\gamma+\nu)}{8+\nu}\right\},
\ee
while for the balance  $_{\nu}\mathcal{B}_{2}$ we find,
\be
\textrm{spec}(_{\nu}\mathcal{K}_{2})=\left\{-1,0,\frac{1+2\gamma}{1+\gamma},
\frac{-4+4\gamma-\nu}{4+4\gamma}\right\}.
\ee
The balance $_{\nu}\mathcal{B}_{2}$ implies that we may escape finite-distance
singularities for $-1<\gamma<-1/2$ and $\nu >-4+4\gamma$ since then the last two
eigenvalues of $_{\nu}\mathcal{K}_{2}$ become negative.
We note though that for these ranges of $\gamma$ and $\nu$ the balance
$_{\nu}\mathcal{B}_{1}$ is also valid and it is such that at least the last eigenvalue
of $_{\nu}\mathcal{K}_{1}$ becomes then positive.
We may exploit this fact by keeping $\gamma$ in the interval $(-1,-1/2)$ and restricting
$\nu$ to fall in $(-4+4\gamma,-6)$. For this choice of parameters,
the last two eigenvalues of $_{\nu}\mathcal{K}_{1}$ have opposite signs which means
that the solution described by $_{\nu}\mathcal{B}_{1}$ is neither valid around a
finite-distance singularity, nor, around a neighborhood of infinity but rather in the limited
area of an annulus failing thus to provide us with any substantial information about the
asymptotics of the dynamical system (\ref{syst1_1_n_s})-(\ref{syst1_4_n_s}).
We are then left with the unique possibility described by the balance
$_{\nu}\mathcal{B}_{2}$ which is the asymptotic expansion of the general solution
of the system in a neighborhood of infinity. Consequently, avoidance of finite-distance
singularities is feasible for $-1<\gamma<-1/2$ and $-4+4\gamma<\nu<-6$.
\section{Conclusions}
We have studied a model consisting of a three-brane embedded in a five-dimensional bulk
filled with a scalar field and an analog to a perfect fluid possessing a general equation of
state $P_{2}=\gamma\rho_{2}$, characterized by the constant parameter $\gamma$. The two
bulk matter components may act independently, or, they may interact with each other by
exchanging energy in a way that the total energy is conserved.

We have started off by analyzing the evolution of our model in the case that the two bulk
matter components do not interact with each other (the two interaction
parameters are set to zero in this case). We have shown that a quite general feature of the
asymptotic behavior of such model is the emergence of a finite-distance
singularity that is of the collapse type I, II or big rip class. The
singularities accommodated in the first two classes share common
characteristics such as the vanishing of the warp factor. However, the derivative
of the warp factor behaves differently in each case: it is divergent in the collapse type I
class whereas it remains finite in the collapse type II class. The
collapse type I singularity may arise for all values of $\gamma$, whereas,
the collapse type II class arises only when $\gamma<-1/2$ as
this is illustrated by the balance $_{III}\mathcal{B}_{4}$ but also by the balances
$_{II}\mathcal{B}_{3}$, $_{I}\mathcal{B}_{3}$ and $_{I}\mathcal{B}_{4}$ for an adequate
choice of their parameters. All of these balances lead to particular solutions
for this range of $\gamma$.
The third class, on the other hand, of big rip singularities, arises always with $\gamma<-1$ and it
is characterized by the divergence of the warp factor, its
derivative and density of the fluid while the energy density of the
scalar field now tends to zero in the neighborhood of the
singularity.

We completed our analysis for non-interacting bulk matter by addressing the
important issue of whether it is possible to avoid finite-distance singularities.
We found that this is true only for a curved brane and for $\gamma>-1/2$.
This is demonstrated by the balances $_{III}\mathcal{B}_{4}$, $_{II}\mathcal{B}_{3}$ and
$_{I}\mathcal{B}_{4}$ that give negative $\mathcal{K}$-exponents
for $\gamma>-1/2$. For a flat brane on the other hand, the avoidance of singularities is not
possible, as this was discussed in detail in the last paragraph of Subsection 3.7.
The main reason for the failure to escape finite-distance singularities in the case of flat brane
is that for all values of $\gamma$ there always exists a balance that corresponds to a
general solution and describes its behavior around finite-distance singularities due to the presence
of the scalar field component in the bulk. It is thus impossible to find ranges of $\gamma$
that they are not characterized by singular behavior.

For illustration, we present a summary of our results for this first analysis in the Table~1 
below, using the notation for the various singularities introduced in Section 2 and the
symbol $\ast$ to denote a balance that corresponds to a particular solution. For each entry
in the table we have taken into account all the corresponding examples in our analysis to deduce the form of each balance and $\mathcal{K}$-exponents.
Note that we used the numerical examples only as representatives of the corresponding asymptotic behaviors for the different regions of the parameter $\gamma$.
We also give the ranges of $r$ and $s$, entering in the definitions of the balances around the type I and II singularities defined in Subsections 3.1 and 3.2,
that lead to the most general behavior possible for each balance.

We continued our analysis to include also cases with interacting bulk matter,
motivated by the fact that it is impossible to find in our flat brane model ranges of
$\gamma$ that lead to avoidance of singularities within finite distance in the case of
non-interacting bulk matter. We studied the case of interaction $\sigma=0$, $\nu$
arbitrary. We have shown that this choice leads to the avoidance of
finite-distance singularities for $-1<\gamma<-1/2$ and $-4+4\gamma<\nu<-6$. For a
curved brane, avoidance of finite-distance singularities is allowed for $\gamma\geq-1/2$,
$\nu>-6$ and $\sigma=0$. These results enforce our previous conclusion about the possibility of such solutions and show that they are also possible in the more involved case considered in this paper, thus proving that the self-tuning mechanism may be sustained under field interactions in the bulk.

We illustrate the results we found for the case of non-interacting bulk components as well as for the case with interaction $\sigma=0$ and $\nu$ arbitrary in the
following tables:
\begin{table}
{\footnotesize
\begin{center}
\begin{tabular}{|c|c|c|c|c|}
\hline
equation of state             &\mco{2}{c|}{flat brane} & \mco{2}{c|}{curved brane} \\
\hline
$P_{2}=\gamma\rho_{2}$ & type                             & balance                                                                                                                    & type                                 & balance
\\ \hline
$\gamma> 1$                  & singular type I                     & $_{II}\mathcal{B}_{1}$, $_{III}\mathcal{B}_{2}$,                                                       & {\em regular}                    & at $\infty$: $_{III}\mathcal{B}_{4}$,
\\
                                        &                                             & $_{II}\mathcal{B}_{2}$ ($-1<r<-2/(\gamma+1)$)                                                        &                                        & $_{I}\mathcal{B}_{4}$ ($-4(\gamma+1)<s<-2$),
\\
                                        &                                             &                                                                                                                                  &                                        & $_{II}\mathcal{B}_{3}$ $(-4<r<-1)$
\\
\hline
$\gamma=1$                   & singular type I                     & $_{III}\mathcal{B}_{1}$                                                          & {\em regular}                    &at $\infty$: $_{III}\mathcal{B}_{4}$,
\\
                                        &                                             &                                                                                                                                &                                         & $_{I}\mathcal{B}_{4}$ ($-8<s<-2$),
\\
                                        &                                             &                                                                                                                                &                                        & $_{II}\mathcal{B}_{3}$ $(-4<r<-1)$
\\
\hline
$-1/2<\gamma<1$          & singular type I             & $_{I}\mathcal{B}_{1}$, $_{III}\mathcal{B}_{3}$,                                                           & {\em regular}               & at $\infty$: $_{III}\mathcal{B}_{4}$,
\\
                                       &                                      & $_{I}\mathcal{B}_{2}$ ($-2<s<-(1+\gamma)$)                                                              &                                     &  $_{I}\mathcal{B}_{4}$ ($-4(\gamma+1)<s<-2$),
\\
                                        &                                     &                                                                                                                                   &                                     & $_{II}\mathcal{B}_{3}$ $(-4<r<-1)$
\\
\hline
$\gamma=-1/2$              &   {\em regular}            & at $\infty$ $_{I}\mathcal{B}_{3}$                                                                 & {\em regular}                                 & at $\infty$: $_{I}\mathcal{B}_{3}$, $_{I}\mathcal{B}_{4}$ ($s>-2$)
\\
                                        &  singular type I            & $_{I}\mathcal{B}_{2}$ ($s<-1/2$), $_{III}\mathcal{B}_{3}$                                                             & singular type I                               &$_{I}\mathcal{B}_{1}$
\\
\hline
$-1<\gamma<-1/2$         & {\em regular}                 & at $\infty$: $_{III}\mathcal{B}_{2}$,                                                                                       & {\em regular}                                & $_{III}\mathcal{B}_{4}^{\ast}$, $_{I}\mathcal{B}_{4}^{\ast}$ ($s<-2$),
\\
                                        &                                      & $_{II}\mathcal{B}_{2}$ ($-2/(\gamma+1)<r<-1$) &                                                                                                       & $_{II}\mathcal{B}_{3}^{\ast}$ ($-4<r<-1$)
\\
                                        & singular type I               & $_{I}\mathcal{B}_{1}$, $_{III}\mathcal{B}_{3}$,           &     singular type II                              & $_{I}\mathcal{B}_{4}^{\ast}$ ($-2<s<-4(1+\gamma)$),
\\
                                        &                                       & $_{I}\mathcal{B}_{2}$ ($-2<s<-(1+\gamma)$)              &                                                           &$_{III}\mathcal{B}_{4}^{\ast}$, $_{I}\mathcal{B}_{3}^{\ast}$, $_{II}\mathcal{B}_{3}^{\ast}$ ($r>-1$)
\\ \hline
$\gamma=-1$                  & singular type I               & $_{I}\mathcal{B}_{1}$, $_{III}\mathcal{B}_{3}$,           &    {\em regular}                                   & at $\infty$: $_{I}\mathcal{B}_{4}^{\ast}$ ($s<-2$), $_{III}\mathcal{B}_{4}^{\ast}$,
\\
                                        &                                       & $_{I}\mathcal{B}_{2}$ ($-2<s<0$)                                 &                                                          &$_{II}\mathcal{B}_{3}^{\ast}$ ($-4<r<-1$)
\\
                                       &                                        &                                                                                   &   singular type II                                & $_{I}\mathcal{B}_{4}^{\ast}$ ($-2<s<-4(1+\gamma)$),
\\
                                       &                                        &                                                                                   &                                                           & $_{III}\mathcal{B}_{4}^{\ast}$, $_{I}\mathcal{B}_{3}^{\ast}$, $_{II}\mathcal{B}_{3}^{\ast}$ ($r>-1$)
\\
\hline
$\gamma<-1$                  & singular big rip               & $_{II}\mathcal{B}_{1}$, $_{III}\mathcal{B}_{2}$,        &    {\em regular}                                   & at $\infty$: $_{I}\mathcal{B}_{4}^{\ast}$ ($s<-2$), $_{III}\mathcal{B}_{4}^{\ast}$,
\\
                                       &                                        & $_{II}\mathcal{B}_{2}$ ($-1<r<-2/(\gamma+1)$)       &                                                            & $_{II}\mathcal{B}_{3}^{\ast}$ ($-4<r<-1$)
\\
                                       &  singular type I                & $_{I}\mathcal{B}_{1}$, $_{III}\mathcal{B}_{3}$,           &      singular type II                               & $_{I}\mathcal{B}_{4}^{\ast}$ ($-2<s<-4(1+\gamma)$),
\\
                                       &                                        & $_{I}\mathcal{B}_{2}$ ($-2<s<-(1+\gamma)$)               &                                                             & $_{III}\mathcal{B}_{4}^{\ast}$, $_{I}\mathcal{B}_{3}^{\ast}$, $_{II}\mathcal{B}_{3}^{\ast}$ ($r>-1$)
\\
\hline
\end{tabular}
\end{center}
}
\label{summary1}
\caption{Summary of our results for the case of non-interacting bulk components.}
\end{table}
\pagebreak
\begin{table}
{\small
\begin{center}
\begin{tabular}{|c|c|c|c|c|}
\hline
equation of state &\mco{2}{c|}{flat brane} & \mco{2}{c|}{curved brane} \\
\hline
$P_{2}=\gamma\rho_{2}$ & type                        & balance                                                                                 & type                    & balance
\\ \hline
$\gamma>1$                   &  {\em regular}        & $_{\nu}\mathcal{B}_{1}$: $-8<\nu<-6$ at $\infty$                 & {\em regular}       & $_{\nu}\mathcal{B}_{3}$: $\nu>-6$ at $\infty$
\\
                                        & singular big rip       & $_{\nu}\mathcal{B}_{1}$: $\nu<-8$                                       &                            &
\\
                                        &   singular type I      & $_{\nu}\mathcal{B}_{2}$: $\nu<-4+4\gamma$                      &                            &
\\
\hline
$\gamma=1$                   & singular type I          &$_{\nu}\mathcal{B}_{2}$: $\nu<0$                                     & {\em regular}      & $_{\nu}\mathcal{B}_{3}$: $\nu>-6$ at $\infty$
\\
\hline
$-1/2<\gamma<1$          & singular type I           & $_{\nu}\mathcal{B}_{2}$: $\nu<-4+4\gamma$,                & {\em regular}    & $_{\nu}\mathcal{B}_{3}$: $\nu>-6$ at $\infty$
\\
                                       &                                   & $_{\nu}\mathcal{B}_{1}$: $\nu>-4+4\gamma$                  &                          &
\\
\hline
$\gamma=-1/2$              & {\em regular}              & $_{\nu}\mathcal{B}_{-1/2}$: $\nu>-6$ at $\infty$                         & {\em regular}  & $_{\nu}\mathcal{B}_{-1/2}$: $\nu>-6$ at $\infty$
\\
                                        & singular type II           & $_{\nu}\mathcal{B}_{-1/2}$: $\nu<-6$                           & singular type II    & $_{\nu}\mathcal{B}_{-1/2}$: $\nu<-6$
\\
                                        &   singular type I           & $_{\nu}\mathcal{B}_{1}$:  $-6<\nu<0$        &                             &
\\
\hline
$-1<\gamma<-1/2$         & {\em regular}                & $_{\nu}\mathcal{B}_{2}$: $-4+4\gamma<\nu<-6$,         & singular type II        & $_{\nu}\mathcal{B}_{3}$: $\nu<-6$
\\
                                        &                                      & $\nu>-6$  at $\infty$                                                   &          &
\\
                                        & singular type I              & $_{\nu}\mathcal{B}_{1}$: $\nu>-6$                               &          &
\\
\hline
$\gamma=-1$                  & singular type I              & $_{\nu}\mathcal{B}_{1}$: $-6<\nu<0$                          & singular type II & $_{\nu}\mathcal{B}_{3}$: $\nu<-6$
\\
\hline
$\gamma<-1$                 & singular big rip              & $_{\nu}\mathcal{B}_{2}$: $\nu>-4+4\gamma$            & singular type II     & $_{\nu}\mathcal{B}_{3}$: $\nu<-6$
\\
                                       & singular type I               & $_{\nu}\mathcal{B}_{1}$: $\nu>-6$                             &      &

\\ \hline
\end{tabular}
\end{center}}
\label{summary2}
\caption{Summary of our results for the case of interacting bulk components.}
\end{table}
\section*{Acknowledgements}
I.K. is grateful to CERN-TH, where part of her work was done, for financial support that made her visits there possible and for allowing her to use its excellent facilities.  The work of I.A. was supported in part by the European Commission under the ERC Advanced Grant 226371 and the contract PITN-GA-2009-237920.
{}
\end{document}